\documentclass[twocolumn,letterpaper,amsmath,amssymb,floatfix,aps,superscriptaddress]{revtex4}

\usepackage{hyperref}
\usepackage{verbatim}
\usepackage{graphicx}
\usepackage{bm} % bold math
\usepackage{ifpdf} % This lets us get figures right both in pdf and ps versions
\usepackage{dcolumn}  % Align table columns on decimal point
\usepackage{epsfig}
\usepackage{color}

\newcommand{\Av}[1]{{\bf #1}}

\newcommand {\vct}[1] {\mathbf {#1}}

\def\ln{{\operatorname{ln}}} 
\def\det{{\operatorname{det}}}

\def\bomega{{\mbox{\boldmath $\omega $}}}

\def\bnabla{{\mbox{\boldmath $\nabla $}}}

\def\rmd{{\mathrm{d}}}
\def\rmi{{\mathrm{i}}}
\def\rme{{\mathrm{e}}}

\begin{document}

%\title{Perspective: Electrostatic interactions -- weak coupling, strong coupling, in between and beyond}

\title{Perspective: Coulomb fluids -- weak coupling, strong coupling, in between and beyond}

\author{Ali Naji}
\affiliation{School of Physics, Institute for Research in Fundamental Sciences (IPM), Tehran 19395-5531, Iran}

\author{Matej Kandu\v c}
\affiliation{Department of Physics, Free University Berlin, D-14195 Berlin, Germany}
\affiliation{Department of Theoretical Physics, J. Stefan Institute, SI-1000 Ljubljana, Slovenia}

\author{Jan Forsman}
\affiliation{Theoretical Chemistry, Chemical Center P.O.Box 124
S-221 00 Lund, Sweden}

\author{Rudolf Podgornik}
\affiliation{Department of Theoretical Physics, J. Stefan Institute, SI-1000 Ljubljana, Slovenia}
\affiliation{Department of Physics, Faculty of Mathematics and Physics, University of Ljubljana, SI-1000 Ljubljana, Slovenia}
\affiliation{Department of Physics, University of Massachusetts, Amherst, MA 01003, USA}

\begin{abstract}
We present a personal view on the current state of statistical mechanics of Coulomb fluids with special emphasis on the
interactions between macromolecular surfaces, concentrating on the {\em weak} and the {\em strong} coupling limits. Both 
are introduced for a (primitive) counterion-only system in the presence of macroscopic, uniformly charged boundaries, where they can 
be derived systematically. Later we show how this formalism can be generalized to the cases with additional characteristic 
length scales that introduce new coupling parameters  into the problem. 
These cases most notably include asymmetric ionic mixtures with mono- and multivalent ions that couple differently to charged surfaces, ions with internal charge (multipolar) structure and finite static polarizability, where weak and strong coupling limits can be constructed by analogy with the counterion-only case and  lead to important new insights into their properties that can not be derived by any other means.
\end{abstract}
\maketitle

\setlength\arraycolsep{2pt}

\section{Introduction} 

A {\em Coulomb fluid}  refers specifically to any mobile (thermalized) collection of charges, which interact via Coulomb interactions. These charges may be small ions (such as Na$^{+}$ and Ca$^{2+}$ with sizes around 0.1 nm),  ions with internal structure (such as the rod-like spermine and spermidine ions with sizes around a nanometer) as well as charged nano-particles and macromolecules (such as proteins, colloids, polymers and membranes), which are typically immersed in an aqueous solvent, as is the case in soft- and bio-matter context \cite{Israelachvili,VO,Hunter,Safranbook,andelman-rev,holm,French-RMP,Baus}.  Macromolecular surfaces get charged mainly due to the dissociation of their surface chemical groups in the solvent; this process releases small ions (referred to as co- and counterions depending on the sign of their charge with respect to that of the surface)  into the solution. The type, magnitude, and the particular distribution of surface charges and those of the resulting mobile ions depend  also  on the specific chemistry of the surface as well as a number of other key factors including, among other things, the $p$H, temperature, dielectric properties or, in general, the specific molecular properties of the solvent. Once dissolved, these mobile charges mediate the interactions between macromolecular surfaces.

The presence of Coulomb fluids can significantly alter electrostatic interactions of charged (bounding) surfaces and modify the behaviour of charged macromolecular dispersions \cite{Israelachvili,VO,Hunter,Safranbook,andelman-rev,holm}. The role of Coulomb fluids can be so drastic that, as evidenced by numerous experimental and theoretical investigations over the last several years  \cite{Khan,Quirk,Bloom,Bloom2,Kekicheff93,Della,Della2,Dubois,Strey98,Tang,Tang03,Tang03b,Angelini03,Wong2010,Guld84,Svensson,Bratko86,Guld86,Wood88,Valleau,Kjellander92,Lyub95,Gron97,Gron98,Wu,AllahyarovPRL,Stevens99,LinsePRL,Linse00,Hribar,Messina00,AndreEPL,AndrePRL,AndreEPJE,Deserno03,Naji_epl04,Naji_epje04,Lee04,Kjellander84,Kjellander84b,Attard-etal,Attard-etal2,Podgornik89,Podgornik89b,Barrat,Pincus98,Podgornik98,Ha,Ha2, Safran99, Kardar99,Netz-orland,Ha01,Lau02,Gonz01,Stevens90,Diehl99,Rouzina96,Korny,Arenzon99,Arenzon99b,Shklovs99,Shklovs02,Diehl01,Lau01,Netz01,Levin02,Rouzina98,Golestan99,Levin99,Shklo00,Andre02,Safran02,hoda_review,Naji_PhysicaA,Ali-rev-Singa,Needleman,Pelta,Yoshikawa1,Yoshikawa2,Plum,Raspaud,Savithri1987,deFrutos2005,Siber,Forsman04,Weeks,Weeks2,trulsson,SCdressed1,SCdressed2,SCdressed3,Naji_CCT,Naji_CCT2,jho-prl,asim,Matej-cyl,Matej-cyl2,kanduc,Hatlo2012,Santangelo,Hatlo-Lue,trizac,trizac2,Burak04,olli,Netz-orland2,Manghi1,Manghi2,Manghi3,Messina_rev,Oosawa,Oosawa_book,patey,Manning}, it can challenge our understanding of electrostatic effects
as evidenced by non-conventional phenomena such as electrostatic attraction between like-charged surfaces, especially when multivalent ions (or counterions) are present in the system. Like-charge attraction is manifested in many experimental examples; a few notable cases include formation of large aggregates of like-charged polymers such as microtubules \cite{Needleman} and F-actin \cite{Angelini03,Tang} 
% Add examples charged membranes 
as well as formation of large condensates of DNA in the bulk \cite{Bloom2,Yoshikawa1,Yoshikawa2,Pelta} and in the DNA packaging inside viral shells that are observed in the presence of multivalent cations \cite{Plum,Raspaud,Savithri1987,deFrutos2005,Siber}.
Similarly, numerous numerical simulations have already demonstrated the emergence of like-charge attraction and investigated its 
underlying mechanism in many examples including charged membranes, colloids and polymers \cite{Guld84,Svensson,Bratko86,Guld86,Wood88,Valleau,Kjellander92,Lyub95,Gron97,Gron98,Wu,AllahyarovPRL,Stevens99,LinsePRL,Linse00,Hribar,Messina00,AndreEPL,AndrePRL,AndreEPJE,Deserno03,Naji_epl04,Naji_epje04,Lee04,Andre02,hoda_review,Naji_PhysicaA,Ali-rev-Singa,Naji_CCT,Naji_CCT2,Weeks,Weeks2,SCdressed1,SCdressed2,SCdressed3,jho-prl,asim,Forsman04,trulsson}.

All major theoretical proposals that aim to explain the phenomenon of like-charge attraction go beyond the standard mean-field or Poisson-Boltzmann (PB) theories, which have been studied since the early years of the last century \cite{Israelachvili,VO,Hunter}, 
by including the effects of electrostatic fluctuations and correlations that are neglected in the description of Coulomb fluids on the mean-field level.  
%\cite{Guld84,Svensson,Bratko86,Guld86,Wood88,Valleau,Kjellander92,Lyub95,Gron97,Gron98,Wu,AllahyarovPRL,Stevens99,LinsePRL,Linse00,Hribar,Messina00,AndreEPL,AndrePRL,AndreEPJE,Deserno03,Naji_epl04,Lee04,Kjellander84,Kjellander84b,Attard-etal,Attard-etal2,Podgornik89,Podgornik89b,Barrat,Pincus98,Podgornik98,Ha,Ha2, Safran99, Kardar99,Netz-orland,Ha01,Lau02,Gonz01,Stevens90,Diehl99,Rouzina96,Korny,Arenzon99,Arenzon99b,Shklovs99,Shklovs02,Diehl01,Lau01,Netz01,Levin02,Naji_epje04,Rouzina98,Golestan99,Levin99,Shklo00,Andre02,Burak04,trulsson}. % add Lue and others
These proposals include integral-equation methods (see, e.g., Refs. \cite{Kjellander92,Kjellander84,Kjellander84b}), perturbative improvement of the mean-field theory including loop expansions and other Gaussian-fluctuations approximations (see, e.g., Refs. \cite{Attard-etal,Attard-etal2,Podgornik89,Podgornik89b,Barrat,Pincus98,Podgornik98,Ha,Ha2, Safran99, Kardar99,Netz-orland,Ha01,Lau02,Safran02}), variational methods (see, e.g., Refs. \cite{Netz-orland2,Manghi1,Manghi2,Manghi3}) and local density functional theory (see, e.g., Refs. \cite{Stevens90,Diehl99}). These approaches turn out to be applicable mostly at large separations between charged surfaces or, generally, for relatively small coupling (and/or electrostatic correlation) strengths. This regime is known as the weak-coupling (WC) or ``high-temperature" regime. A complementary strong-coupling (SC) or ``low-temperature" approach was pioneered by Rouzina and Bloomfield \cite{Rouzina96} based on the observation that counterions tend to form two-dimensional strongly-correlated layers at oppositely charged bounding surfaces  when the coupling parameter is large. Such structural correlations can lead to dominant attractive forces of mainly energetic origin between like-charged surfaces that have been studied by means of several different theoretical approaches in recent years (see, e.g., Refs.\cite{AndrePRL,AndreEPJE,Andre02,Naji_epl04,Naji_epje04,Arenzon99,Arenzon99b,Shklovs99,Shklovs02,Diehl01,Lau01,Netz01,Levin02,hoda_review,Naji_PhysicaA,Ali-rev-Singa,trizac,trizac2,Santangelo,Weeks,Weeks2,Hatlo-Lue,Forsman04,Burak04}).
 
For a Coulomb fluid consisting of only a single charge species next to oppositely charged boundaries (the so-called ``counterion-only" model), the WC and SC coupling theories were shown to follow systematically as two limiting laws of a single unified formalism \cite{Netz01},  a view that was completely corroborated by extensive numerical simulations (see, e.g., Refs. \cite{AndreEPL,AndrePRL,AndreEPJE,Naji_epl04,Naji_epje04,Andre02,hoda_review,Naji_PhysicaA,Ali-rev-Singa,Naji_CCT,Naji_CCT2,SCdressed1,SCdressed2,SCdressed3,jho-prl,Weeks,Weeks2,asim}) or available exact solutions \cite{exact1,exact2}. 

In what follows, we provide a guided personalized tour of recent advances in the theory of Coulomb fluids by reviewing various aspects of the WC-SC paradigm within the primitive counterion-only model. We then discuss how this framework can be generalized to derive powerful limiting laws for more complicated, but also more realistic models of Coulomb fluids with additional characteristic length scales (coupling parameters) than envisioned in the original framework of the WC-SC dichotomy; most notably, asymmetric ionic mixtures with mono- and multivalent ions, ions with internal multipolar structure and finite static polarizability next to charged bounding surfaces (see Fig.~\ref{schematic}).

\section{Coulomb fluids: General considerations}

\subsection{The primitive model}

\begin{figure}
 \begin{center}
\includegraphics[angle=0,width=\linewidth]{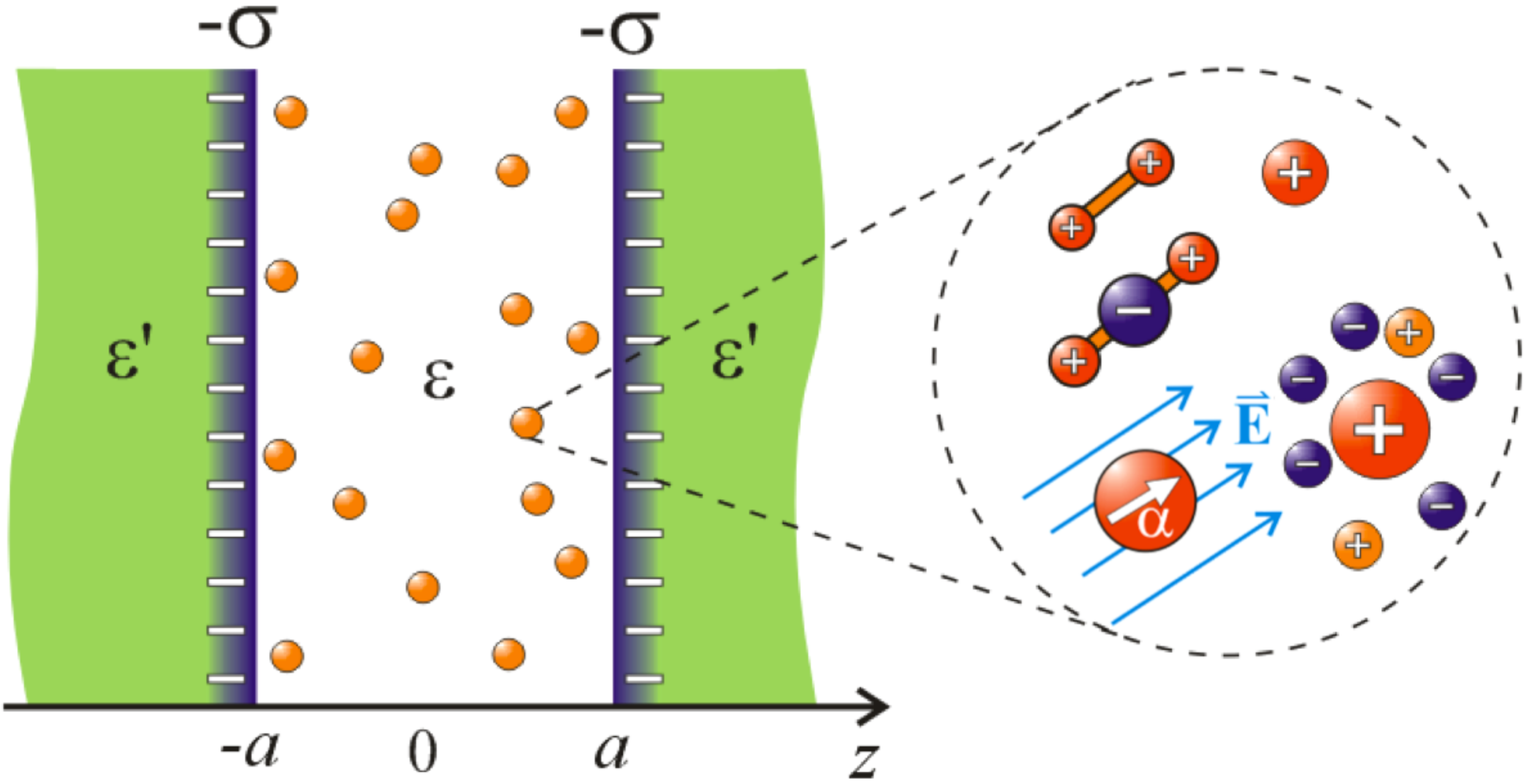}
\end{center}
\caption{Schematic representation of some of possible generalizations of the primitive ``counterion-only" model as considered in this paper. It shows a system of large (mobile) multivalent ions (shown by red spheres), which may possess an internal structure (e.g., a rod-like shape or a static polarizability) and may be dispersed in a bathing solution of monovalent anions and cations in the background (shown by blue and orange spheres, respectively).}
\label{schematic}
\end{figure}

The stated features of Coulomb fluids make a full understanding of the equilibrium properties of 
charged macromolecular systems quite difficult and many aspects of real systems have been either neglected (depending on the particular case or the application under investigation) or heavily approximated, often though with reasonable justification. 
Perhaps the most simple and yet efficient idealization is the so-called {\em primitive model} where solvent is treated  as a featureless continuum of a fixed dielectric constant $\varepsilon$ stemming from the degrees of freedom associated with the solvent molecules; small mobile ions are taken to be featureless and characterized only by their charge, without any internal structure and/or polarizability. Furthermore, macromolecules or other macroscopic surfaces are often treated as objects with a fixed uniform charge distribution (see Fig.~\ref{schematic}). 

A great deal of theoretical effort has been devoted recently to improve upon these simplifying assumptions, in particular regarding the solvent structure (see, e.g., Refs. \cite{Israelachvili,holm,benyaakov,Burak_solvent,Burak_solvent2,Ben-Yaakov2011,Ben-Yaakov2011b}  and references therein) and the polarizability and internal structure of mobile ions as conceived by Debye \cite{Debye} (see, e.g., Refs. \cite{Bikerman,multipoles,Demery, Hatlo,Netz-polar,Horinek}  and references therein), etc. Unfortunately, many of these interesting developments such as ionic liquids \cite{Ionicliquids,IL2,IL3}, surface ion-adsorption effects \cite{Belloni95,Forsman06}, the discreteness and/or heterogeneity of the surface charge distribution (see, e.g., Refs. \cite{andelman-disorder,kantor-disorder1,klein,Meyer,Ben-Yaakov-dis,netz-disorder,ali-rudi,rudiali,partial,disorder-PRL,jcp2010,pre2011,epje2012,jcp2012} and reference therein), $p$H-controlled charge regulation \cite{Regulation,Regulation2,Regulation3,Olvera,Olvera2}, many-body interactions \cite{jure1,jure2} or Bjerrum pair formation \cite{Bjerrumpairing1,Bjerrumpairing2,Bjerrumpairing3} must remain outside the domain of this review. Even on the level of the above simplified model assumptions, the underlying physics of the Coulomb fluids exhibits conceptual challenges and intriguing results, which have only recently been corroborated by computer simulations and, less often, experimental evidence.

%(except for ions with multipolar structure and static polarizability that we shall discuss later,  Fig.~\ref{schematic}) and many other fascinating topics such as ionic liquids \cite{Ionicliquids,IL2,IL3}, surface ion-adsorption effects \cite{Belloni95,Forsman06}, the discreteness and/or heterogeneity of the surface charge distribution (see, e.g., Refs. \cite{andelman-disorder,kantor-disorder1,klein,Meyer,Ben-Yaakov-dis,netz-disorder,ali-rudi,rudiali,partial,disorder-PRL,jcp2010,pre2011,epje2012,jcp2012} and reference therein), $p$H-controlled charge regulation \cite{Regulation,Regulation2,Regulation3,Olvera,Olvera2}, or Bjerrum pair formation \cite{Bjerrumpairing1,Bjerrumpairing2,Bjerrumpairing3} must remain outside of this review. Even on the level of the above simplified model assumptions, the underlying physics of the Coulomb fluids exhibits conceptual challenges and intriguing results, which have only recently been corroborated by computer simulations and, less often, experimental evidence.
	
\begin{figure*}[t!]
\includegraphics[width=15cm]{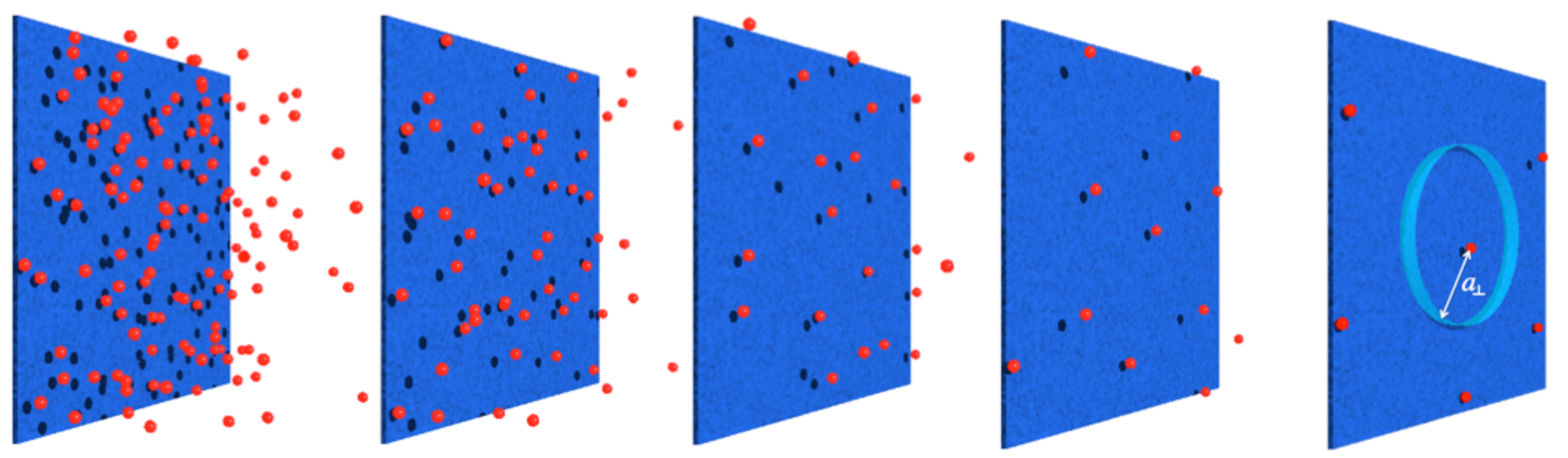}
\caption{An extended 3D counterion layer in the WC regime,  $\Xi\ll 1$, reduces to a quasi-2D layer in the SC regime, $\Xi\gg 1$, upon increasing the coupling parameter (from left to right). In the limit of $\Xi \rightarrow \infty$, a single-particle description becomes relevant because of large correlation holes, $a_{\perp} \rightarrow \infty$, forming around individual counterions.}
\label{fig:xi}
\end{figure*}	

\subsection{Physical scales and parameters}
\label{subsec:scales}

Of the plethora of possible length scales (associated with electric charges, van der Waals (vdW) or chemical bonding energies, ion sizes, solvent molecular size, or spacing between discrete surface charges, etc., see Ref. \cite{French-RMP}), one can argue that within the primitive model discussed above only a few will be important.

The thermal energy $k_{\mathrm{B}}T = \beta^{-1}$ can be compared with inter-ionic Coulomb interactions (e.g., for two elementary charges $e_0$), giving rise to the Bjerrum length $\ell_{\mathrm{B}}= e_0^2/(4\pi \varepsilon \varepsilon_0 k_{\mathrm{B}}T)$,
with gas-like behavior for small and liquid- or solid-like behavior for large $\ell_{\mathrm{B}}$. For an ionic fluid consisting of ions (counterions) of charge valency $q$ at a charged interface of uniform surface charge density $-\sigma e_0$, 
the Gouy-Chapman length $\mu=1/(2\pi  q \ell_{\mathrm{B}} |\sigma|)$ determines the strength of the thermal energy with respect to the  electrostatic interaction (attraction) with the surface (we assume that $q>0$ and $-\sigma<0$ with no loss of generality).  The neutralizing counterions next to a charged surface are thus expected to form a diffuse gas-like phase when the {\em electrostatic coupling parameter}, defined as the ratio
\begin{equation}
\Xi = {q^2 \ell_{\mathrm{B}}}/{\mu} = 2\pi q^3\ell_{\mathrm{B}}^2 |\sigma|,
\label{aclcadfls}
\end{equation}
is small, i.e., $\Xi\ll 1$. In the opposite situation with large $\Xi\gg 1$, one observes a very different behavior \cite{Netz01,AndreEPL,AndrePRL,AndreEPJE,hoda_review,Naji_PhysicaA,Ali-rev-Singa,Shklovs99,Shklovs02,Diehl01,Lau01,Levin02,trizac,trizac2}, see Fig.~\ref{fig:xi}, where the ionic cloud is reduced from a three-dimensional (3D) layer to a quasi-two-dimensional (2D) sheet. For $\Xi\gg 1$, the mean separation between counterions, which follows from the local electroneutrality condition as $a_\bot\sim \sqrt{q/|\sigma|} \sim \mu \sqrt{\Xi}$, is much bigger than the thickness of the surface (counterion) layer given by $\mu$.
%and hence the ratio 
%\begin{equation}
%\frac{a_\bot}{\mu} \sim \left(\frac{q^2\ell_{\mathrm{B}}}{\mu}\right)^{1/2} =\sqrt{\Xi}. 
%\label{eq:abot}
%\end{equation}
Conjointly with the surface-counterion interaction, the inter-counterion repulsions, $q^2\ell_{\mathrm{B}}/a_\bot \sim \sqrt{\Xi}$, also dominate the thermal energy. Therefore, increasing $\Xi$ engenders a strongly-correlated liquid, or even a crystalline phase of counterions, in the limit $\Xi \rightarrow \infty$. 

This picture follows from extensive Monte-Carlo (MC) simulations \cite{AndreEPL,AndrePRL,AndreEPJE,Andre02,hoda_review,Naji_PhysicaA}, which show that the {\em crossover} from the {\em weak-coupling (WC) regime} ($\Xi\ll 1$) to the {\em strong-coupling (SC) regime} ($\Xi\gg 1$) is associated with a hump in the heat capacity of the system 
%(due to reduction in the dimensionality of the counterion layer as noted above) 
and development of short-range correlations between counterions in the range $10<\Xi<100$. The transition to a Wigner crystalline phase (characterized by a diverging heat capacity) then occurs at a very large value of the coupling parameter around $\Xi\simeq 3\times 10^4$ \cite{AndreEPL,AndrePRL,AndreEPJE,Baus}. 

The WC-SC paradigm was derived from a functional-integral (field-theoretic) representation of the partition function directly and systematically only in the case of the primitive counterion-only model   \cite{Netz01}. The presence of additional mobile charge components, surface charge heterogeneity or mobile ion multipolar structure, and/or polarizability introduces new length scales and thus new coupling parameters into the problem. Nevertheless, one can always identify a principal coupling parameter, in analogy to the counterion-only case, that implies a SC-like fixed point. The thusly defined SC fixed point then displays a fine structure and can bifurcate into other strongly coupled states governed by these additional length scales and the ensuing coupling parameters.  In what follows, we will 
%show a personalized selection of 
discuss a few illuminating examples (see Fig.~\ref{schematic}) displaying the salient features of this approach and its main results.

\section{Field theory and the WC-SC paradigm in the counterion-only model}

The preceding observations suggest that the WC and the SC regimes should follow as complementary descriptions from a single, unified formalism. This clearly transpires from the field-theoretic approach to Coulomb fluids \cite{Edwards,Podgornik89,Podgornik89b,Netz01,Netz-orland}, wherein the WC and the SC limits can indeed be derived as {\em asymptotic} theories from a single {\em ansatz} for $\Xi\rightarrow 0$ and $\Xi\rightarrow \infty$, respectively \cite{Netz01}. 

In general, the Hubbard-Stratonovich transformation allows the partition function of the primitive model to be mapped {\em exactly} to a functional integral over a fluctuating (electrostatic) potential, $\phi({\mathbf{r}})$  \cite{Edwards,Podgornik89,Podgornik89b,Netz01,Netz-orland}, as
\begin{equation}
  {\mathcal Z} = \int \!{\mathcal D}\phi \,\, \rme^{-\beta S[\phi]}, 
\end{equation}
where the effective ``field-action" can be written (up to a ``self-energy" prefactor, which we ignore here, see, e.g.,  \cite{Podgornik89,Podgornik89b,Netz01,Netz-orland,SCdressed1,SCdressed2,SCdressed3})
as
\begin{widetext}
\begin{equation}
\label{fieldaction}
  S[\phi] =  \frac{1}{2} \iint \rmd {\mathbf{r}}  \rmd {\mathbf{r}'} \phi(\mathbf{r}) u^{-1}({\mathbf{r}}, {\mathbf{r}}') \phi(\mathbf{r}') + \rmi \int \rmd {\mathbf{r}}\,\rho_0({\mathbf{r}})\phi({\mathbf{r}}) -  \int \rmd {\mathbf{r}}~{\mathcal U}(\phi(\mathbf{r}))  = S_0[\phi(\mathbf{r})] -  \int \rmd {\mathbf{r}}~{\mathcal U}(\phi(\mathbf{r})),
\end{equation}
\end{widetext}
where the Coulomb interaction is $u({\mathbf{r}}, {\mathbf{r}}')  = 1/(4\pi~\varepsilon\varepsilon_0 |{\mathbf{r}} -{\mathbf{r}}'|)$ and its kernel (operator inverse) $u^{-1}({\mathbf{r}}, {\mathbf{r}}')  = -\varepsilon_0 \nabla\cdot \varepsilon({\mathbf r}) \nabla \delta({\mathbf{r}} -{\mathbf{r}}')$. The fixed, external (macromolecular) charge distribution is described by $\rho_0({\mathbf{r}})$ and is assumed, for further specification, to be distributed on planar surfaces with either one plate (placed at $z=0$) or two plane-parallel surfaces (placed at $z=-a$ and $z = +a$, i.e., at separation distance $D=2a$ along the $z$ axis) with a uniform surface charge density of $-\sigma e_0$. In addition, the system can exhibit a nontrivial spatial dielectric constant profile $\varepsilon({\mathbf{r}})$. The model of the Coulomb fluid is embodied in the generally non-linear ``field self-interaction" term ${\mathcal U}(\phi(\mathbf{r}))$.  The presence of these non-linear field self-interactions renders an exact evaluation of the partition function difficult, except in a few special cases (for such exact solutions, see, e.g.,   Refs. \cite{exact1,exact2} and references therein). Nonetheless, remarkable analytical progress has been made in analyzing the behavior of such systems, especially %in the presence of multivalent ions,  which 
in the cases that go beyond the usual WC-SC framework (see, e.g., Refs. \cite{Netz01,SCdressed1,SCdressed2,SCdressed3,multipoles,Demery}).

%In what follows we focus on a few simple yet illuminating model systems. In all cases, 
%\begin{equation}
%\label{eq:sigma}
%  \rho({\mathbf{r}}) = \left\{
%	\begin{array}{ll}
%	 \sigma \delta(z) & \quad {\textrm{one plate}}\\
%	 \sigma [\delta(z-a)+\delta(z+a)] & \quad {\textrm{two plates}}
%	\end{array}
%\right.
%\end{equation}

In the {\em counterion-only model}, one assumes that only counterions of charge valency $q$ are present in the system and exactly neutralize the fixed charges.  
%In most realistic situations though, there is always a variable amount of salt in the system, especially,  when multivalent counterions are present. 
Although this model may be inapplicable in most real situations, it has nevertheless served as a useful paradigm, elucidating fundamental  aspects of the WC-SC dichotomy.  In this context, the field self-interaction is obtained as \cite{Podgornik89,Podgornik89b,Netz01}
\begin{equation}
  {\mathcal U}(\phi(\mathbf{r})) = \lambda \Omega({\mathbf{r}}) \rme^{-\rmi \beta q e_0\phi } \equiv \lambda\,  \tilde{\mathcal U}(\phi(\mathbf{r})).
  \label{eq:V}
\end{equation}
Here, $\lambda$ is the  fugacity of counterions and  the ``blip" function $ \Omega({\mathbf{r}})$ defines the region of space allowed to the mobile counterions (i.e., it is equal to one in the region where counterions are allowed to be present and zero elsewhere). The functional integral can be cast in a dimensionless form that depends on only one parameter, i.e., $\Xi$ \cite{Netz01}. The WC-SC limits then follow directly from appropriate evaluations of the partition function \cite{Netz01}.

\subsection{WC theory: Mean-field theory and weak fluctuations}

The {\em WC limit}, $\Xi \rightarrow 0$, for this model is obtained by solving the saddle-point equation of the field-action  \cite{Podgornik89,Podgornik89b,Netz01}, 
\begin{equation}
\frac{\delta S[\phi]}{\delta \phi}\bigg\vert_{\rmi \phi=\psi_{\mathrm{PB}}} = 0,
\label{eq:var_S}
\end{equation}
which governs the {\em mean} (real-valued) electrostatic potential $\psi_{\mathrm{PB}}({\mathbf{r}}) = \rmi \langle \phi({\mathbf{r}})\rangle$ and leads to the standard Poisson-Boltzmann (PB) equation  
\begin{equation}
- \varepsilon_0\nabla \cdot \left[ \varepsilon({\mathbf{r}}) \nabla \psi_{\mathrm{PB}}({\mathbf{r}}) \right]= \rho_0({\mathbf{r}}) + q e_0 \lambda\, \Omega({\mathbf{r}}) \, \rme^{-\beta q e_0 \psi_{\mathrm{PB}}({\mathbf{r}})},
  \label{eq:PB}
\end{equation}
with Neumann boundary conditions at the charged interfaces.  For one or two planar surfaces, the dielectric discontinuity at the bounding surfaces plays no role and the above equation leads to the standard Gouy-Chapman theory of electrical double layers \cite{Israelachvili,VO,Hunter,Safranbook,andelman-rev,holm} with $\psi_{\mathrm{PB}}({\mathbf{r}}) = \psi_{\mathrm{PB}}(z)$.  The corresponding expressions for the counterion density profile $n_{\mathrm{PB}}(z)$ and the disjoining pressure $P_{\mathrm{PB}}$ imparted by the counterions on the bounding surfaces are well known \cite{Safranbook,andelman-rev}, with the latter obtained from the contact-value theorem \cite{contact_value,contact_value2,DeanCT} through the counterion density at the mid-plane $z=0$. This leads to the following asymptotic form for large $D = 2a$,
\begin{equation}
P_{\mathrm{PB}}(D) = k_{\mathrm{B}}T n_{\mathrm{PB}}(z; D)\bigg|_{z=0} \simeq  \frac{k_{\mathrm{B}}T}{q^2\ell_{\mathrm{B}} D^2}\bigg|_{D \rightarrow \infty},
\end{equation}
which clearly shows that the PB pressure at large separations is independent of $\sigma $, has an entropic origin and is thus repulsive. Quite generally, it can be proven that the disjoining pressure within the PB theory is always {\em repulsive}, regardless of the shape of the charged surfaces as long as the boundary conditions are symmetric \cite{Neu,Sader,Sader2}. 

The PB equation is exact in the strict limit of $\Xi\rightarrow 0$ and has been applied successfully to study weakly charged systems \cite{Safranbook,andelman-rev,VO,Israelachvili,Hunter}.  When $\Xi$ is finite but small, one expects subdominant Gaussian fluctuations to occur around the PB solution, $ \phi = -\rmi \psi_{\mathrm{PB}} + \varphi$. These fluctuations are described by the field-action
\begin{eqnarray}
&&\!\!S[\phi] \simeq S[-\rmi \psi_{\mathrm{PB}}] +\\
&&\!\!\qquad +\frac{1}{2} \int {\mathrm{d}}{\mathbf r}\,{\mathrm{d}}{\mathbf r}'\,  \varphi({\vct r}) \varphi({\vct r}')\, \frac{\delta^2 S[\phi]}{\delta \phi({\vct r}) \delta \phi({\vct r}')}\bigg\vert_{\phi=-\rmi \psi_{\mathrm{PB}}} \!\!\!\!+ {\cal O}(\varphi^3),\nonumber
\end{eqnarray}
and lead to small deviations from the PB pressure. The total WC pressure can then be written as 
\begin{equation}
P_{\mathrm{WC}}(D)= P_{\mathrm{PB}}(D) + \Xi \, P_1(D) + {\mathcal O}(\Xi^2), 
\end{equation}
where $P_1$ is the Gaussian (one-loop) correction around the mean-field solution, analogous to vdW-type interactions  \cite{Podgornik89,Podgornik89b,Netz01,Netz-orland,Parsegian2005} and is thus {\em attractive}, $P_1<0$. For two like-charged surfaces, 
\begin{equation}
\Xi \, P_1(D) \sim -k_{\mathrm{B}}T\,\frac{\ln D}{D^3}\bigg|_{D \rightarrow \infty}.
\end{equation}
which has a similar algebraic dependence on $D$ as in the case of the classical vdW force between neutral dielectrics, generated
by the thermal fluctuations of the zero-frequency Matsubara modes of the electromagnetic field \cite{Parsegian2005}; in the present case, however, 
the attractive correction is caused by the Gaussian fluctuations around a {\em non-uniform background} defined by the 
PB solution that produces also a logarithmic $D$-dependent correction. 
It should be noted that in the WC regime the fluctuation-induced attraction can not overcome the repulsive leading-order PB pressure, the total pressure thus remaining {\em repulsive}. 

In certain models that assume surface condensation or adsorption of counterions on (fixed) charged boundaries \cite{Ha,Ha2,Pincus98,Kardar99,Ha01,Lau02,Safran99,Safran02}, the repulsive mean-field effects are strongly suppressed and the total pressure can eventually turn out to be attractive. Such condensation behavior can not occur within the primitive model with purely Coulombic interactions and may occur only if one takes into account non-electrostatic surface effects going beyond the models considered here. 

\subsection{SC theory}

As shown by Netz \cite{Netz01}, a systematic description for the {\em SC limit} follows from the same field-theoretic formalism that yields the PB equation in the limit $\Xi\rightarrow 0$, provided that one takes the limit of $\Xi\rightarrow \infty$. In this case, one can perform a fugacity (virial) expansion combined with a $1/\Xi$ expansion, the leading order of which was shown to be finite and given by {\em single-particle} contributions only. Physically, since in this limit the counterions are strongly attracted to the surface and are isolated in large correlation holes (of size $\sim a_\bot$) surrounding them, the partition function of the system is expected to be dominated by single-particle contributions from the interaction between counterions and charged surfaces. To the lowest order, one finds 
\begin{eqnarray}
   {\mathcal Z}_{\rm SC} = %&& \int \!{\mathcal D}\phi \, \rme^{-\beta S_0[\phi]} \left( 1 +  \lambda \int \rmd {\mathbf{r}}~\tilde{\mathcal U}(\phi)  + {\cal O}({\lambda}^2)\right) \qquad  
\label{partfun} 
 %\nonumber\\
  &&  {\mathcal Z}_{\rm SC}^{(0)} + \lambda  {\mathcal Z}_{\rm SC}^{(1)} + {\cal O}(\lambda^2)
    \label{partfun1},
\end{eqnarray}
%This was show to lead to a well-defined limiting SC theory, which is formally obtained by first virial-expanding the partition function in powers of the fugacity of counterions and then expanding it in powers of $1/\Xi$. Unlike in the case of bulk electrolytes, the virial expansion in the presence of fixed, charge macroscopic  objects is well-defined and converges to finite results for thermodynamic quantities. 
%To the leading order the partition function Eq. (\ref{partfun}) has a single-particle form involving only the interaction between counterions and charged surfaces as the dominant form of the interactions in the system, assuming the form
%\begin{eqnarray}
%  {\mathcal Z}_{\rm SC} \simeq  {\mathcal Z}_{\rm SC}^{(0)} + \lambda  {\mathcal Z}_{\rm SC}^{(1)} + {\cal O}(\lambda^2) \nonumber\\
%    {\mathcal Z}_{\rm SC}^{(0)} \left(1+ \lambda \int e^{-\beta e_0 q \int u({\vct r}, {\vct r}') \rho({\vct r}) d^3{\vct r}} d^3{\vct r}' \right), 
%  \label{partfun1}
%\end{eqnarray}
where the first term %in Eq.~(\ref{partfun1}) 
involves only  $S_0[\phi(\vct r)]$ of (\ref{fieldaction}) due to external charges, while
\begin{equation}
{{\mathcal Z}_{\rm SC}^{(1)}}/{{\mathcal Z}_{\rm SC}^{(0)}} = \int  \rmd{\vct r}\,\Omega({\mathbf{r}}) \, \rme^{-\beta e_0 q \int  \rmd{\vct r'} \, u({\vct r}, {\vct r}') \rho_0({\vct r}')}, 
  \label{partfun2}
\end{equation}
is the single-ion partition function in the field of external charges (note the grand-canonical formalism, which can be transformed back to a canonical description involving $N$ particles by a Legendre transform \cite{Netz01}).  As discussed in Sec. \ref{subsec:scales}, counterion-counterion interactions are also very strong in the limit $\Xi\rightarrow \infty$, but appear as subleading if compared with counterion-surface contributions. 

%In the case of planar bounding surfaces, only degrees of freedom in the $z$ direction perpendicular to the surfaces should determine  the behavior of the system.  The leading order term in the above expansion is then expected to  dominate in the regime of distances $z<a_\bot$ from a charged surface, or equivalently $z/\mu < \sqrt{\Xi}$ (this argument needs to be amended in the case where dielectric inhomogeneities are present in the system, which we neglect throughout out discussion, see Refs.  \cite{AndreEPL,AndrePRL,AndreEPJE,SCdressed2,SCdressed3} and references therein). 

In the SC limit and for two like-charged bounding surfaces, the disjoining pressure can be expressed to the leading-order as \cite{Netz01}
\begin{equation}
%P(D)= P_{\mathrm{SC}}(D) + \Xi^{-1} \, P_1(D) + {\mathcal O}(\Xi^{-2}), 
%\end{equation}
%where the term reads  \cite{Netz01}
%\begin{equation}
P_{\mathrm{SC}}(D) = -\frac{\sigma^2}{2\varepsilon\varepsilon_0}+\bigg(\frac{2|\sigma|}{q}\bigg)\frac{k_{\mathrm{B}}T}{D}.
\label{eq:PSC}
\end{equation}
It is easy to see that the first term follows from the electrostatic energy of individual counterions sandwiched between two equally charged surfaces, leading to an {\em attractive} pressure, while the {\em repulsive} second term is clearly nothing but the ideal-gas entropy of the counterions confined to the slit between the surfaces. In fact, as noted above in the SC limit, the counterions are isolated in correlation holes of size $a_{\perp}$ so that $ |\sigma| a_{\perp}^2 \sim q$, and hence the energetic repulsion between the two like-charged surfaces is overcompensated by the attraction between them and the individual counterion in between in such a way that it changes the sign of the pressure  \cite{Naji_PhysicaA,hoda_review,Netz01}! 

The above result holds for small separations $D< a_{\perp}$ or equivalently, $D/\mu < \sqrt{\Xi}$. Within the regime of validity of the SC limit, one recovers both the repulsive regime for $D<D_*$ and the attractive regime for $D>D_* $, where $D_*=2\mu$ is the separation at which the total force acting on the surfaces vanishes, as confirmed by MC simulations \cite{AndreEPL,AndrePRL,AndreEPJE} (this argument needs to be amended in the case where dielectric inhomogeneities are present in the system, see, e.g., Refs. \cite{AndreEPL,AndrePRL,AndreEPJE,SCdressed2,SCdressed3} and references therein). 

\begin{figure*}[t]\begin{center}
\begin{minipage}[b]{0.35\textwidth}\begin{center}
\includegraphics[width=\textwidth]{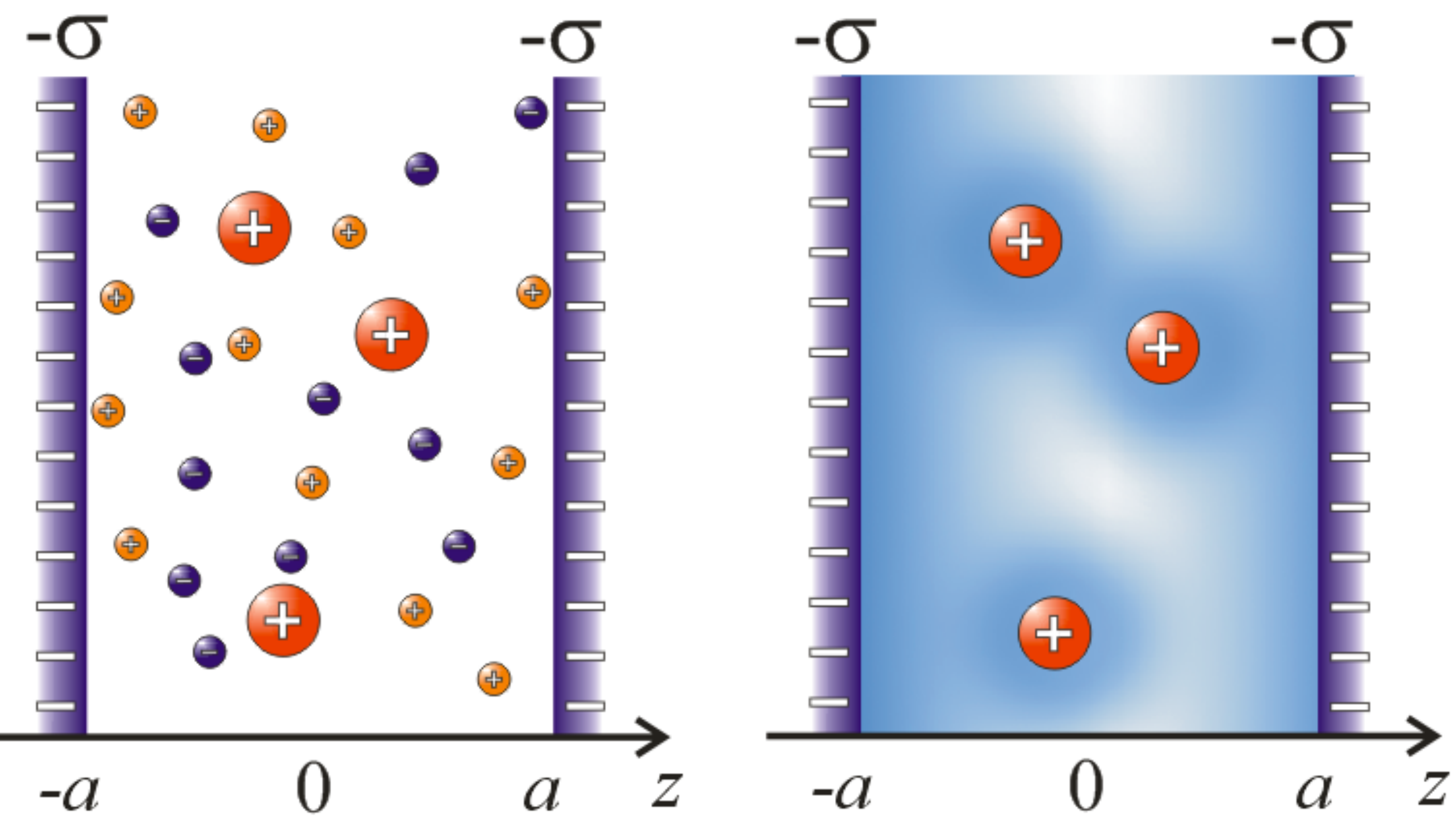}\\ \phantom{x} (a)
\end{center}\end{minipage}\hskip0.1cm
\begin{minipage}[b]{0.31\textwidth}\begin{center}
\includegraphics[width=\textwidth]{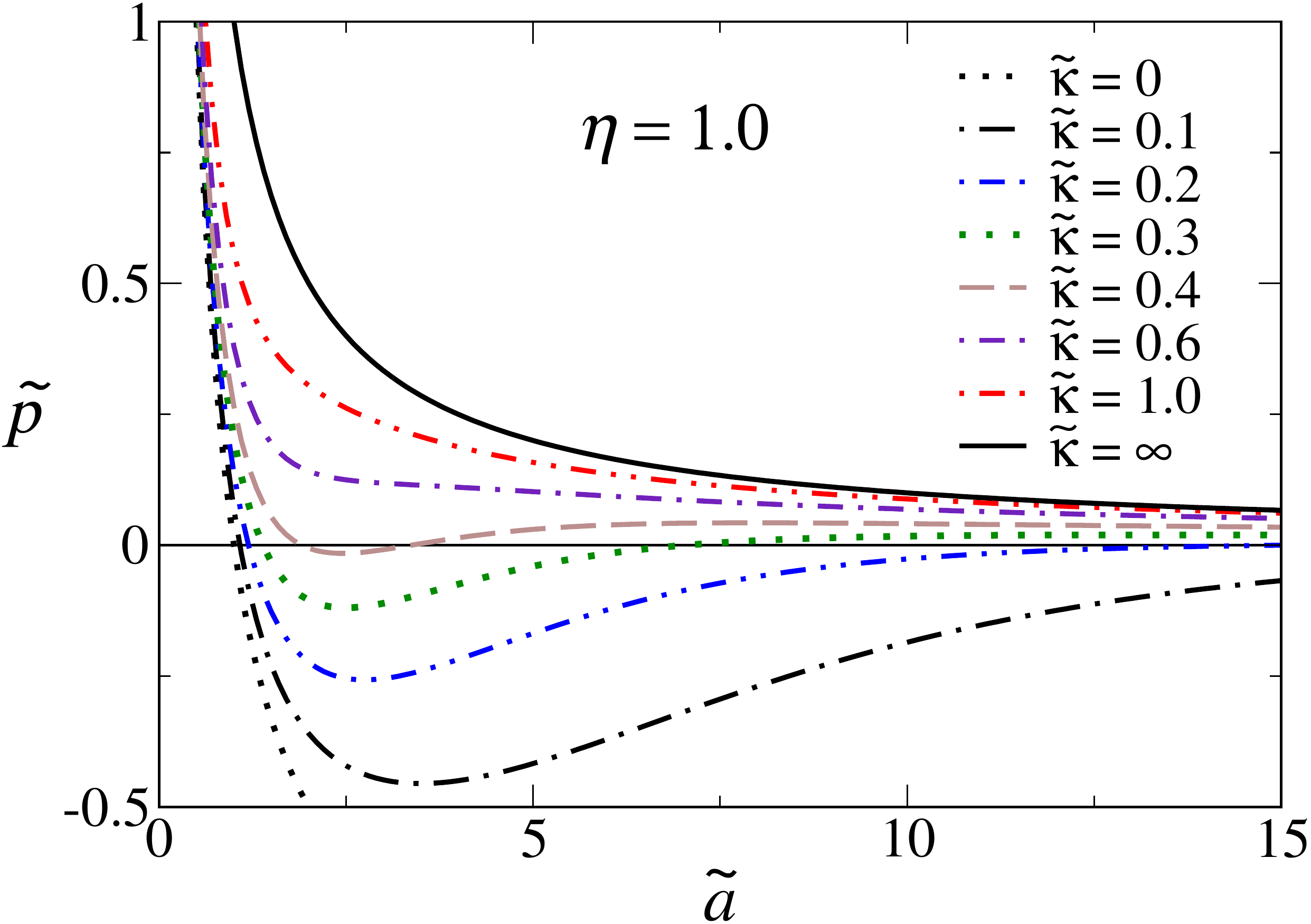} (b)
\end{center}\end{minipage}\hskip0.1cm
\begin{minipage}[b]{0.31\textwidth}\begin{center}
\includegraphics[width=\textwidth]{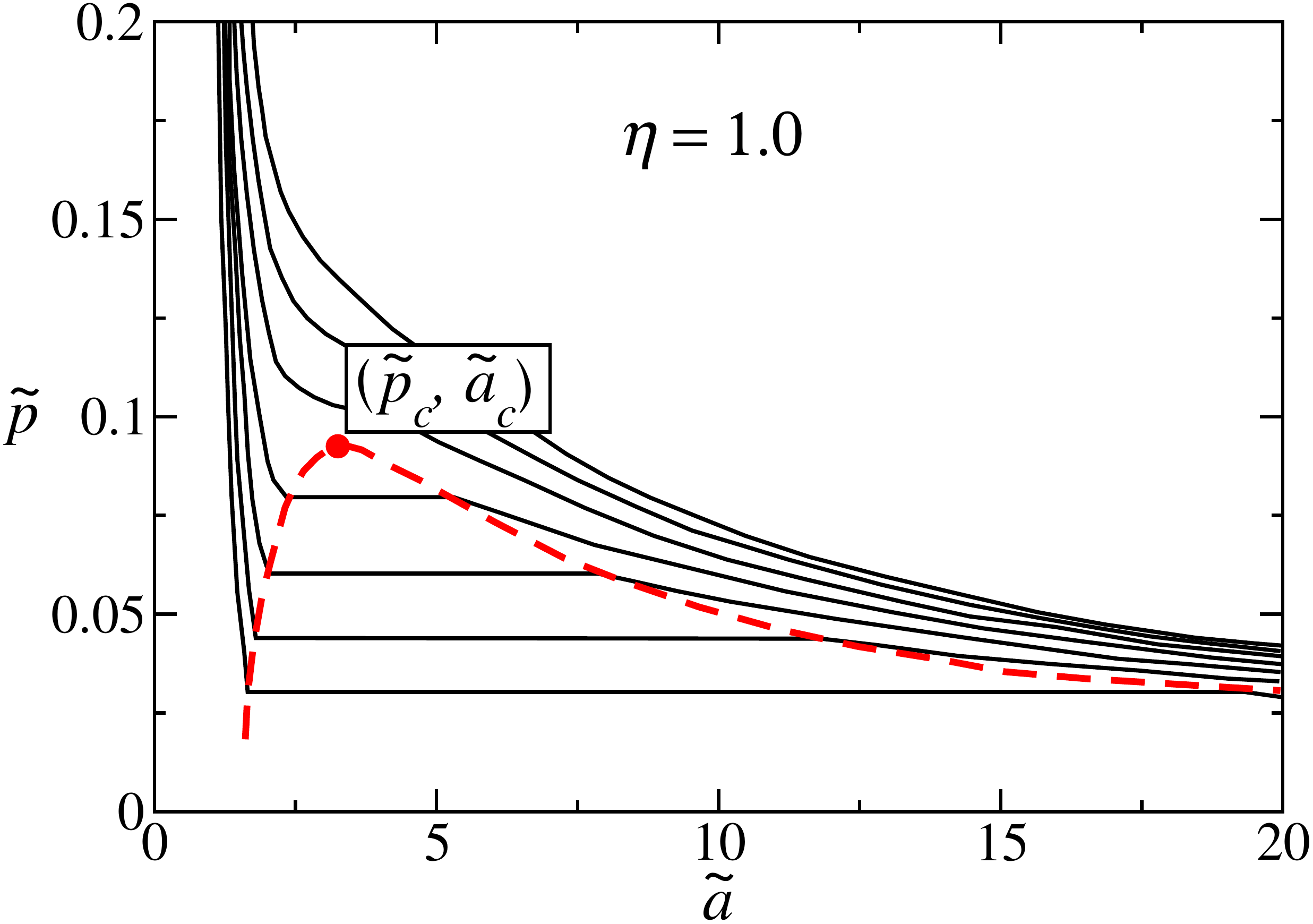} (c)
\end{center}\end{minipage}\hskip0.1cm
\caption{(a) A mixture of multivalent  counterions (red spheres)  in a bathing solution of  monovalent salt ions (blue and orange spheres); the former are strongly and the latter are weakly coupled to external (surface) charges. The monovalent ions can be integrated out from the explicit microscopic model (left) in the case of a highly asymmetric mixture (with highly charged multivalent ions), giving rise to the effective dressed multivalent-ion model (right). (b) The rescaled interaction pressure, $\tilde p=\beta P_{\textrm{SC}}/2\pi \ell_{\mathrm{B}}\sigma^2$, as predicted by the dressed multivalent-ion theory, Eq.~(\ref{p_C}), 
is shown as a fucntion
of the rescaled half-distance between the surfaces, $\tilde a=a/\mu$. It bridges between the pure osmotic regime of multivalent counterions (large $\kappa$) where the electrostatic interactions are screened out and the SC theory for a counterion-only system (small $\kappa$), which exhibits attractive interaction between the like-charged surfaces for a wide range of parameters (here shown
for $\eta=1$, i.e., when the total charge due to multivalent counterions exactly compensates the surface charges).  
(c) In the canonical dressed multivalent-ion ensemble, there appears a non-monotonic behavior for the  interaction pressure showing vdW-like loops, suggesting a coexistence regime between two different phases. The phase coexistence region actually ends with a critical point ($p_c, a_c$). Adapted from Ref. \cite{SCdressed1}.}
\label{fig:asymmetric}
\end{center}\end{figure*}

\subsection{Crossover from WC to SC limit}

For separation distances larger than the correlation hole size or at smaller coupling parameters, one needs to account for the higher-order terms in the SC expansion. Even though the single-particle SC limit appears to work very well for coupling parameters down to $\Xi\sim 100$ (corresponding, e.g., to DNA with trivalent counterions), the perturbative expansion turns out to be very inefficient and can hardly be extended into the so-called {\em crossover regime}, i.e., $10<\Xi<100$ \cite{Naji_PhysicaA,hoda_review}. This regime remains accessible {\em fully} only through simulations. 
%As noted before, the crossover regime corresponds to the regime of parameter where  the counterion layer becomes quasi-two-dimensional and exhibits a pronounced correlation hole and liquid-like oscillations in the pair distribution function of counterions \cite{AndreEPL,AndrePRL,AndreEPJE}.  
Despite the challenging nature of correlations in this regime, Santangelo  \cite{Santangelo}, Weeks et al. \cite{Weeks,Weeks2} and others \cite{Hatlo-Lue,Forsman04} have shown that approximations based on the {\em decomposition method} used within the liquid-state theories can be applied with reasonable success. Also a test-charge theory developed by Burak et al. \cite{Burak04} seems to capture the onset of correlations and provide very useful insights into the crossover behavior of the counterion-only system. 

Another approach to study counterion-only systems in the limit of large couplings has been proposed by considering the ground-state structure of the system in the limit $T\rightarrow 0$ \cite{Rouzina96,Shklovs02}. It should be noted that this latter limit does not in general coincide with the SC limit, which is based on a combined virial and $1/\Xi$ expansion.
% which entails taking the limit $\Xi\rightarrow \infty$. 
The virial approach leads to finite entropic (temperature) corrections at the leading order (e.g., the second term in Eq.~(\ref{eq:PSC})) and deviations have been reported for the subleading corrections calculated from ground-state considerations and those obtained from the virial approach \cite{trizac,trizac2}.   

We should also note that {\em modified PB equations} are sometimes invoked as {\em ad hoc} generalizations of the standard PB theory that include the ion excluded-volume term and the so-called fluctuation potential \cite{MPB,MPB2}. While this kind of theories certainly describe some features of steric and correlation interactions inaccessible to the standard PB theory, they can not be viewed as a limiting law in the sense that it follows systematically from a single unified formalism valid in both the weak and the strong coupling limit.

\section{Generalizations of the WC-SC paradigm}

In the original form \cite{AndreEPJE,Netz01,hoda_review,Naji_PhysicaA}, the WC and SC dichotomy applies only to a counterion-only system confined by uniformly charged surfaces, a situation of profound theoretical significance but seldom encountered in the real world. While the WC limit, through its correspondence with the saddle-point equation, can be defined for any field-action, the existence of additional length scales, besides the Bjerrum and the Gouy-Chapman ones,  precludes a direct introduction of a unique electrostatic coupling parameter and consequently a systematic derivation of the SC limit with its virial expansion.

In this case, however,  one can generalize not the SC theory itself, but the single-particle aspect of the SC limit, which can be then applied to any system that contains highly charged species, irrespective of the number of governing length scales (see, e.g., Refs. \cite{SCdressed1,SCdressed2,multipoles,Demery,Naji_epl04,Naji_epje04,Naji_CCT,Naji_CCT2,Matej-cyl2,ali-rudi}). The approach is reasonable because the highly charged ions are usually present only in small concentrations and thus a virial expansion, of which the single-particle limit is the lowest order, makes sense. In addition, the single-particle SC limit has been tested against detailed simulations and its regime of applicability is determined for any particular case. 

In quite general context, for any field-action, one could thus derive a saddle-point theory as a substitute for the WC limit and a single-particle theory as a substitute for the SC limit.
% (where both these theory should reproduce the standard WC and SC theories when the appropriate limit in the parameter space is taken). 
This is a powerful consequence of the field-theoretic approach and can be applied even to the cases where other approaches  (such as those based on the ground-state methods \cite{trizac,trizac2}) are not necessarily applicable (e.g., asymmetric ionic mixtures containing large amounts of monovalent salt). This is what we will illustrate for a few interesting cases detailed below.

\subsection{Asymmetric ionic mixtures: Dressed multivalent-ion approach}

\begin{figure}[t!]
\includegraphics[width=5cm]{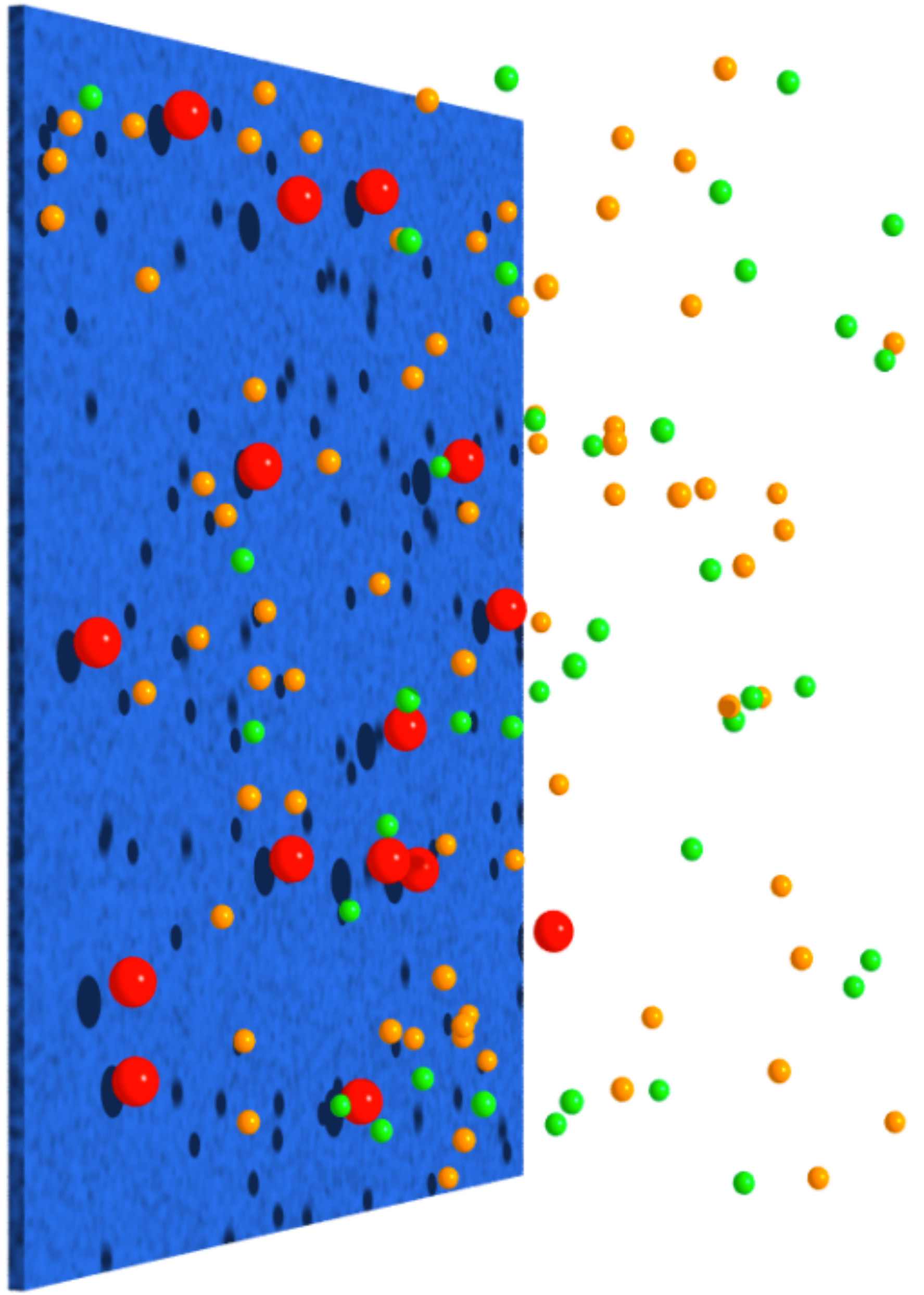}
\caption{Snapshot of an explicit MC simulation of a single (negatively) charged surface  
with (positive) tetravalent counterions (red spheres) and monovalent salt anions and cations (green and orange spheres); 
while the former are strongly attracted toward the surface, the latter form an extended DH-like atmosphere. (In this case, there is also a dielectric discontinuity at the charged surface corresponding to that  of  the hydrocarbon/water interface).
%; also, for the sake of illustration, sphere radii are enlarged by a factor of 2.5 as compared with the actual hard-sphere radii used in the simulations). 
}
\label{snapshot}
\end{figure}

A particularly relevant case is that of a mixture of {\em multivalent} ions in a bathing solution of {\em monovalent} ions (Fig.~\ref{fig:asymmetric}a and Fig.~\ref{snapshot}). This is a typical situation in the formation of liquid crystalline mesophases of semiflexible biopolymers \cite{rau-1,rau-2,Angelini03}, multivalent-ion driven condensation of DNA in the bulk \cite{Bloom2,Yoshikawa1,Yoshikawa2,Pelta,Plum,Raspaud}, or in viruses, where multivalent ions are believed to play a key role in the stability of the viral capsid and/or packaging of its genome \cite{Savithri1987,deFrutos2005,Siber}. 

For an ionic mixture consisting of a single species of multivalent counterions ($c$) in a neutralizing background of monovalent anions ($-$) and cations ($+$), we have 
\begin{equation}
 {\mathcal U}(\phi({\mathbf{r}})) = \Omega({\mathbf{r}}) \left( \lambda_c  \rme^{-\rmi\beta q e_0  \phi} +\lambda_+ \rme^{-\rmi\beta e_0  \phi} + \lambda_-  \rme^{\rmi  \beta e_0  \phi}\right), 
  \label{eq:V_s}
\end{equation}
which follows directly from the Hubbard-Stratonovich transformation of the  microscopic (Coulomb) Hamiltonian of the system \cite{Podgornik89,Podgornik89b,Netz01,SCdressed1}. Here, $\lambda_c$ and $\lambda_{\pm}$ are the respective ionic fugacities, and the ``blip" function $\Omega({\mathbf{r}})$ is assumed to be the same for all mobile species.  

Obviously, in an asymmetric mixture,  multivalent counterions and monovalent ions are coupled to macromolecular charges quite  differently: multivalents {\em strongly}, while monovalents only {\em weakly}, as evident from their respective electrostatic coupling parameters. This presents a challenging problem as no single limit, neither WC nor SC will apply to all of the components of the system \cite{SCdressed1}. The saving grace in this situation is the fact that usually multivalent ions are present at very small concentrations, e.g., around just a few mM (see, e.g., Refs. \cite{rau-1,rau-2,Pelta,Plum,Raspaud}), and thus their behavior is expected to fit naturally within the virial scheme based on an expansion in terms of their fugacity (bulk concentration).

Additionally, in highly asymmetric solutions with $q\gg1$ (e.g., with tri- and tetravalent counterions), the problem can be furthermore greatly simplified by employing the following approximation \cite{SCdressed1},
\begin{equation}
{\mathcal U}(\phi({\mathbf{r}})) \simeq  \lambda_c \, \rme^{-\rmi \beta q e_0  \phi} -  n_{\mathrm {b}}  (\beta e_0  \phi)^2/2 + {\cal O}(\phi^3)
\end{equation}
(in the region where the ions are present $\Omega({\mathbf{r}}) = 1$), where $n_\textrm  b=2n_0+qc_0$, assuming that the multivalent counterions result from a $q$:1 salt of  bulk concentration $c_0$, which is mixed with a 1:1 salt of bulk concentration $n_0$; hence, $\lambda_c = c_0$,  $\lambda_+=n_0$ and $\lambda_- = n_0+qc_0$ (note that the multivalent salt contributes an additional amount of monovalent ions with concentration $qc_0$). 

From this form, it is clear that the degrees of freedom due to monovalent ions can be integrated out, resulting in an effective formalism including only screened interactions between the remaining  ``dressed'' multivalent ions and fixed macromolecular charges \cite{SCdressed1,SCdressed2,SCdressed3}. This follows from the first term of the field-action $S_0[\phi(\mathbf{r})]$, Eq.~(\ref{fieldaction}), and the quadratic term in the above expansion. In other words, the effective field-action of the system is given by 
%\begin{widetext}
\begin{eqnarray}
%  S[\phi] \simeq && \frac{1}{2} \int \rmd {\mathbf{r}}\rmd {\mathbf{r}'}\phi({\mathbf{r}})  u_{\mathrm{DH}}^{-1}({\mathbf{r}}, {\mathbf{r}}')  \phi({\mathbf{r}}') + i \int \rmd {\mathbf{r}} \, \rho({\mathbf{r}})\phi({\mathbf{r}}) + \lambda_c \int \rmd {\mathbf{r}} \, \Omega_c({\mathbf{r}}) e^{-i q e_0 \beta \phi},
S[\phi] \simeq && S_\textrm{DH}[\phi] + \lambda_c \int \rmd {\mathbf{r}} \, \Omega({\mathbf{r}})\, \rme^{-\rmi \beta q e_0  \phi},
  \label{bhfgjosd}
\end{eqnarray}
%\end{widetext}
where $S_\textrm{DH}[\phi]$ has the same form as $S_0[\phi]$ except that $u^{-1}({\mathbf{r}}, {\mathbf{r}}')$ is replaced by the screened Debye-H\" uckel (DH) kernel $u_{\mathrm{DH}}^{-1}({\mathbf{r}}, {\mathbf{r}}')  = -\varepsilon_0[ \nabla\cdot \varepsilon({\mathbf r}) \nabla - \varepsilon({\mathbf r})\kappa^2] \delta({\mathbf{r}} -{\mathbf{r}}')$, and the inverse Debye length is introduced as $\kappa^2 = 4\pi \ell_{\mathrm{B}} n_\textrm b$. Obviously, the multivalent ion concentration $c_0$ introduces a new parameter and new physics that goes beyond the simple DH screening picture. 
%In the case when $q$ and/or $c_0$ are large enough, 
The multivalent ion effects can be codified by a new length scale defined as 
$ \chi^2 = 8\pi q^2 \ell_{\mathrm{B}} c_0$  \cite{SCdressed1,SCdressed2,SCdressed3}, which is analogous to the Debye length in the case of monovalent salt.
%In the opposite regime with $c_0 > n_0/q^2$, the many-body effects due to multivalent ions dominate and the behavior of the system is explored by means of MC simulations \cite{SCdressed1,SCdressed2,SCdressed3}. 

We refer to the  theory  based on the field-action (\ref{bhfgjosd}) as the {\em dressed multivalent-ion theory}. The key point here is thus that the above approach can be applied only to highly asymmetric ionic mixture with $q\gg 1$ \cite{SCdressed1}. The regime of validity of this approach can be checked against explicit-ion MC simulations, where all ions, including the monovalent ones, are explicitly simulated (see Fig.~\ref{snapshot}), which show that this approach can indeed give quantitatively accurate results in a wide range of realistic parameter values \cite{SCdressed2,SCdressed3}. This may not seem obvious at first since the above approach treats the monovalent ions on the implicit DH level and one might thus expect strong deviations to occur when multivalent ions are present in the solution due, e.g., to nonlinear charge renormalization and/or Bjerrum pairing effects \cite{Bjerrumpairing1,Bjerrumpairing2,Bjerrumpairing3}; these effects, however, turn out to be absent or negligible
in the regime of parameters that is of concern to our discussion \cite{SCdressed1,SCdressed2,SCdressed3}. 

On the saddle-point level of the dressed multivalent-ion field-action, the PB equation, Eq.~(\ref{eq:var_S}), reads 
\begin{equation}
-\varepsilon_0 \left[\nabla \cdot\varepsilon({\mathbf r}) \nabla - \varepsilon({\mathbf r})\kappa^2\right]\psi_{\mathrm{PB}}  = \rho_0 + q e_0 c_0\, \Omega({\vct r})\,  \rme^{-\beta q e_0  \psi_{\mathrm{PB}}}.
\end{equation}
The corresponding disjoining pressure
%with $n_{c,\pm}$ as the respective densities, is given by 
%\begin{equation}
%P_{\mathrm{PB}}(D) = k_{\mathrm{B}}T \bigg( \sum_{i= c, \pm} n_i(z; D)\bigg|_{z=0} - 2 n_0 - c_0\bigg), 
%\end{equation}
%proportional to $\bigg( \sum_{i= c, \pm} n_i(z=0; D)- 2 n_0 - c_0\bigg)$ 
is again consistently repulsive for symmetric boundary conditions. The underlying assumption here is that electrostatic correlations are zero and
thus all ions remain weakly coupled to each other and the external fixed charges.

When $ c_0 < {n_0}/{q^2}$, one can furthermore develop a simple theory based on a virial expansion, which can capture the behavior of such asymmetric solutions in the limit corresponding to a generalized SC limit, where multivalent counterions are coupled strongly (to each other and the external charges), while monovalent ions still remain weakly coupled to other charge species. By inspection of the field-action (\ref{bhfgjosd}), the single-particle limit
%, which now takes on the role of the SC theory, 
has exactly the same form as Eq.~(\ref{partfun1}), except that the screened DH potential replaces the Coulomb interaction, i.e., $u({\mathbf{r}}, {\mathbf{r}}') \rightarrow u_\textrm{DH}({\mathbf{r}}, {\mathbf{r}}')$.  For a single charged surface, the dressed multivalent-ion approach can successfully predict the density profile of ions next to the surface as well as the charge inversion induced by multivalent counterions  \cite{SCdressed2}. 

For two apposed like-charged planar surfaces \cite{SCdressed1}, the interaction pressure behaves very differently if the dressed multivalent counterions are treated on the canonical (they are not in chemical equilibrium with a bulk reservoir) or grand-canonical level (in chemical equilibrium with the bulk) \cite{SCdressed1}.
The corresponding interaction pressure (within the canonical ensemble for dressed multivalent counterions and assuming, for the sake of simplicity, that the system is dielectrically homogeneous and that the monovalent salt ions are present both inside and outside the slit) is then transformed from Eq.~(\ref{eq:PSC}) to 
\begin{equation}
P_\textrm{SC}(D)= \frac{(e_0\sigma)^2}{2\varepsilon\varepsilon_0} \left[  \rme^{-\kappa D}+2\mu\eta\, {I'( D)}/{I( D)}\right],
\label{p_C}
\end{equation}
where $\eta= {(N q)}/{2|\sigma| S}$ is the amount of multivalent  counterions in the slit between the surfaces relative to the total surface charge and $I(D=2a)=\int_{-a}^{ a}\exp\Bigl( ({2}/{\kappa\mu})\,\rme^{-\kappa a}\cosh\,\kappa z\Bigr)\,\rmd z $. The case $\eta=0$ represents a system with salt only and the DH theory is recovered, whereas $\eta=1$ is the case when the total charge due to counterions exactly compensates the surface charges; the counterion-only SC theory discussed above can be recovered within this latter case by letting $\kappa\rightarrow 0$, see Fig.~\ref{fig:asymmetric}b. 
%If the multivalent ions are treated within a grand-canonical ensemble of bulk concentration $c_0$, then one finds
%\begin{equation}
%P_\textrm{SC}(D)= \frac{(e_0\sigma^2)}{2\varepsilon\varepsilon_0} \left[  \rme^{-\kappa D}+\chi^2\mu^2\, (I'( D) -1)/4 \right].
%\label{p_GC}
%\end{equation}
This again  illustrates that the dressed multivalent-ion theory can bridge between the standard WC and SC limits. 
For multivalent dressed counterions within the canonical description, the interaction pressure can even become non-monotonic, displaying repulsive branches at small and large inter-surface separations \cite{SCdressed2}, see Fig.~\ref{fig:asymmetric}b and c. 
In fact, for certain values of the parameters, the interaction pressure shows vdW-like loops, which could suggest a coexistence regime between two different phases, see Fig.~\ref{fig:asymmetric}c. The binodal or the coexistence curve actually ends with a critical point corresponding to the largest amount of salt in the system that still leads to a non-monotonic dependence of the pressure on the inter-surface separation.

The dressed multivalent-ion approach can also give an insight into attractive image-induced depletion forces between two {\em neutral} dielectric walls  \cite{SCdressed3}. 
These forces arise as a result of expulsion of multivalent ions from the vicinity of the walls in the slit region in the situation where the two dielectric walls have a dielectric constants smaller than that of water in the slit, as is typically the case. In this case, the dielectric image charges of individual ions have the same sign as the ions themselves and so are repelled from one another, pushing the mobile ions from the slit region back to the bulk solution. Subsequently, this effect leads to a lower osmotic pressure acting on the walls from the ionic solution in the slit, and hence eventually a net inter-surface attraction between the walls due to the bulk osmotic pressure that pushes the surfaces together.  Such depletion-induced forces become important at small inter-surface separations, where they can  even be comparable to or stronger than the usual vdW forces as confirmed also by explicit-ion simulations \cite{SCdressed3}. Similar attractive interactions  can appear in the absence of dielectric discontinuities, albeit they are smaller; namely, in the close proximity of the walls, the multivalent ions are stripped off their ``counterion clouds", leading to an increase of the excess chemical potential,
and thus a decrease of the ideal part (i.e., the density of multivalent ions). 

The polarization depletion mechanism \cite{Monica,Sahin1}, such as the one described above,  shows a generic similarity with the depletion-induced effects in polymer solutions \cite{lekker}. In the present case, however, one should note that the depletion is a consequence of strong correlations between individual ions and their image charges, which turn out to be the dominant leading-order effect in the case of neutral surfaces with multivalent ions, and thus, as such, can be described analytically within the dressed multivalent-ion approach (see Ref. \cite{SCdressed3} for more details).

%The dressed multivalent-ion approach can also give insight into image-induced depletion forces between two dielectric walls \cite{SCdressed3}. The latter arises as a result of {\em polarization depletion} \cite{Monica,Sahin1} of multivalent ions from the slit between two walls with dielectric constants smaller than that of water in the slit, as is typically the case. Dielectric images of individual ions thus have the same sign as the ions themselves and so are repelled  from one another, pushing the mobile ions from the slit region back to the bulk solution.  For multivalent ions in the slit, this effect is quite strong and leads to attractive forces even between {\em neutral} surfaces, which are comparable or stronger that the usual vdW forces and agree well with explicit-ion simulations \cite{SCdressed3}.  

\subsection{Ions with internal charge structure}

\begin{figure}[t!]
\includegraphics[width=7cm]{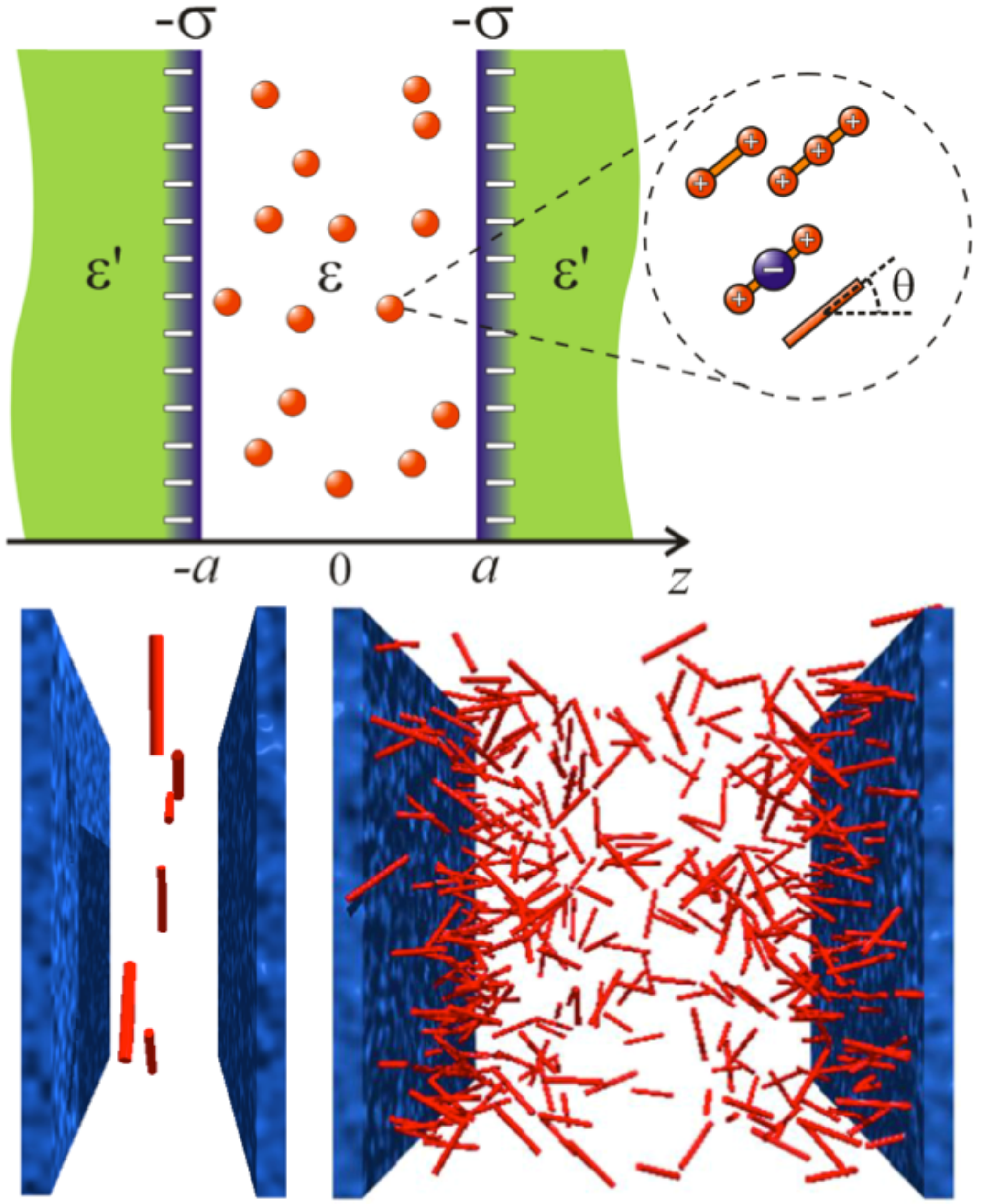}
\caption{Counterions can have an internal structure, such as a rod-like quadrupolar charge distribution (top). For rod-like counterions with non-zero monopolar and quadrupolar moments, the orientational order parameter in the WC limit (bottom, right) indicates preferred orientation perpendicular to the bounding surfaces, whereas in the SC limit (bottom, left) the counterions are preferably aligned parallel to the bounding surfaces. }
\label{fig:rods}
\end{figure}

Multivalent mobile ions are not always describable by simple point-like monopolar charge structure \cite{multipoles}. Some structured counterions can be described with a dipolar \cite{abrashkin,bohinc1} or quadrupolar \cite{bohinc,woon} charge distribution, which introduce additional features in electrostatic interactions that are quite distinct from the standard PB framework. The rigid internal structure of a single counterion can be described by a charge distribution $\hat \rho(\Av r; {\Av R}, \bomega)$, where  ${\Av R}$ gives the location of the counterion  and $\bomega$ are the orientational variables specifying the angular dependence of the counterion charge distribution. Here, we shall focus on the case of uniaxial counterions corresponding, e.g., to mobile charged particles of rod-like structure such as spermine or spermidine.  In this case, the charge distribution $\hat \rho(\Av r; {\Av R}, \bomega)$ can be expanded in a standard multipolar series
%\begin{widetext}
\begin{eqnarray}
&&\hat \rho(\Av r; {\Av R}, \bomega) = e_0q \delta(\Av r - {{\Av R}}) - p_0 (\Av n \cdot \bnabla)\delta(\Av r - {{\Av R}})+\nonumber\\
&&\qquad\qquad\quad + \, t_0 (\Av n \cdot \bnabla)^2 \delta(\Av r - {{\Av R}}) + \cdots,
\label{chaden}
\end{eqnarray}
%\end{widetext}
where $e_0 q$ is the monopolar moment of each counterion, $\Av p = p_0\, \Av n$ is its dipolar moment, and $\Av Q = t_0 \, \Av n \otimes \Av n$ its uniaxial quadrupolar moment with director unit vector $\Av n$.  One possible example where only the monopolar and quadrupolar moments are non-zero is a uniformly charged rod with total charge $e_0 q$ and length $l$ for which $t_0= e_0 q l^2/24$. Another possible case is a rod-like configuration with a negative charge ($-e_2$) in the center and two positive charges ($+e_1$) located at both ends of the rod, giving $e_0 q=2e_1-e_2$ and $t_0=e_1 l^2$, respectively (see Fig.~\ref{fig:rods}, top).

Assuming for simplicity that the system is again composed of counterions only and that these mobile counterions have a complicated internal charge distribution $\hat \rho(\Av r; {\Av R}, \bomega)$, the field self-interaction   in the partition function (see Eq.~(\ref{fieldaction})) can be  derived as \cite{multipoles}  
\begin{equation}
 {\mathcal U}(\phi({\mathbf{r}}))={\lambda}\!\int\!  \rmd \bomega \, \Omega({\vct r})\,\exp\Bigl(\,\rmi\beta \!\!\int \!\! \rmd\Av r'\hat \rho(\Av r'; \Av r, \bomega)\phi(\Av r') \Bigr),
\label{action}
\end{equation}
where $\lambda$ is the fugacity of counterions. The WC limit is obtained from the  saddle point approximation and the SC limit by the lowest-order virial expansion corresponding to a single-particle limit. Though in this case the field-action contains more length scales than one, corresponding to the  internal structure of the counterions with in general $q$, $p_0$, and $t_0$ all having distinct values, one can identify a principal coupling parameter, in analogy to the primitive  counterion-only model, that implies a SC-like fixed point with a fine structure as a consequence of these additional length scales. 

The WC limit for a system of rod-like counterions bounded by two like-charged plane-parallel surfaces is then given by a generalization of the PB equation that 
%depends only on the transverse coordinate $z$, and 
can be written in dimensionless form as \cite{multipoles}
\begin{equation}
\psi'' = -\frac12 {\textstyle}\int_{-1}^{+1}\!\!\!\!\!\!\rmd x ~\Omega(x, z)\left( u(z) - p x u'(z) + t x^2 u''(z) \right),
\label{quadPB1}
\end{equation}
where $\psi$ is the dimensionless potential (i.e., actual potential multiplied by $\beta e_0 q$), $u(z) = {C \exp{\left(- \psi - p x \psi' - tx^2 \psi''\right)}}$ and dimensionless multipolar moments $p=p_0/e_0q\mu$ and $t=t_0/e_0q\mu^2$. Here, $u(z)$ is proportional to the local orientationally-dependent number density of the counterions. 
% and $\rho(z)$ is the corresponding orientationally averaged charge density of the counterions. 
%The characteristic function $\Omega(x, z)$ for the parallel plane geometry simply excludes the counterion configurations that would penetrate the bounding walls and depends on the shape of the counterions. 
Since we are not interested in steric effects, we can set the blip function to $\Omega(x, z) = 1$ everywhere inside the slit between the two surfaces. Furthermore, $x = \cos{\theta}$ is an orientational variable,  where $\theta$ corresponds to the angle between the $z$-axis and the director ${\mathbf n}$.
%the director of the uni-axial counterion is $\Av n$ 
The integral over this variable gives the orientational average. The constant $C$ is set by the boundary conditions on the two bounding surfaces. 

Numerical solutions of Eq.~(\ref{quadPB1}) show that increase of the quadrupolar moment leads to a {\em higher} concentration of counterions at the surfaces  \cite{multipoles}. This can be explained by invoking the potential energy of every quadrupolar particle in electrostatic potential, proportional to the second derivative of the mean-field potential, which in the symmetric case is a concave function of the coordinate $z$. Thus the quadrupolar force acts away from the center toward both surfaces. Furthermore, the orientational order parameter, $S_2={\textstyle\frac{1}{2}}\left(3 \langle x^2\rangle -1\right)$, increases with increasing quadrupolar strength and is larger at both surfaces than at the center, indicating that counterion axes are preferentially aligned parallel to the $z$-axis and thus {\em perpendicular} to the bounding surfaces in the WC limit (see Fig.~\ref{fig:rods}, bottom, right).

%In the case of higher multipoles the number density and the charge density of the counterions are not proportional. Expressions $\rho_i(z)$ are simply the orientationally averaged multipolar charge densities, i.e. $i=1$ for monopolar charge, $i = 2$ for dipolar charge etc. 

The SC limit is  obtained by a virial expansion up to the term linear in $\lambda$, corresponding to a single-particle partition function \cite{multipoles}. One can show that ${\mathcal Z}_{\rm SC}^{(0)}$ is again given by the first term in Eq.~(\ref{partfun}) involving only
%\begin{equation}
%Z_{\rm SC}^{(0)} = \exp\left[- {\textstyle \frac12} \beta \!\!\int \rmd\Av r \rmd\Av r' \rho_0(\Av r) u(\Av r,\Av r') \rho_0(\Av r') \right],
%\label{funint-SC-0}
%\end{equation}
electrostatic interactions between both charged surfaces, so that the multipolar nature of the mobile ions is not important for this lowest-order term. It becomes, however, important at the next order, linear in $\lambda$, which equals to
\begin{widetext}
\begin{equation}
{{\cal Z}_{\rm SC}^{(1)}}/{{\cal Z}_{\rm SC}^{(0)}} =  \!\!\int\!\!\!\!\int \rmd\Av R\, \rmd\bomega\, \Omega({\mathbf r})\,\exp\biggl[- \beta\!\! \int\!\!\!\!\int\!\! \rmd\Av r \rmd\Av r' \hat \rho(\Av r; {\Av R}, \bomega) u(\Av r,\Av r') \rho_0(\Av r') - {\textstyle\frac12} \beta \int\!\!\!\!\int \rmd\Av r \rmd\Av r' \hat \rho(\Av r; {\Av R}, \bomega) u(\Av r,\Av r') \hat \rho(\Av r'; {\Av R}, \bomega)\biggr].
\label{funint-SC-1}
\end{equation}
\end{widetext}
The interaction potential in this part of the partition function is in general composed of the direct and image electrostatic interactions, i.e., $u(\Av r,\Av r') = u_0(\Av r,\Av r') + u_{\mathrm{im}}(\Av r,\Av r')$, if the bounding surfaces have a different dielectric permittivity from the solvent. The second term in the exponent of Eq.~(\ref{funint-SC-1}) gives the image self-interactions among monopolar, dipolar, and quadrupolar moments (9 terms), whose explicit forms have been calculated in Ref. \cite{multipoles}. Note that, in general, the self-energy contributions can not be simply renormalized away into the rescaled fugacity \cite{Netz-orland,selfenergy2}.

In the case of two charged plane-parallel surfaces with uniform surface charge density, the electrostatic potential does not depend on the $z$ coordinate, is thus spatially homogeneous and given by ${\sigma e_0 a}/{(\varepsilon\varepsilon_0)}$.
%\begin{equation}
%\int \rmd\Av r' u(\Av r,\Av r')\rho_0(\Av r')=-{\sigma e_0 a}/{(\varepsilon\varepsilon_0)}$.
%\label{funint-u-00}
%\end{equation}
Since all the terms in the density operator  (\ref{chaden}), except the first one, depend on spatial gradients (derivatives), the counterion energy in a homogenous external electrostatic potential depends only on the monopolar (first) term. The corresponding energy of a counterion in this electrostatic potential, i.e., the first term in the exponent of (\ref{funint-SC-1}) is then given by $ \beta q{\sigma e_0^2 a}/({\varepsilon\varepsilon_0})$.
As for the self-energy term (second term in the exponent of (\ref{funint-SC-1})), it only picks up constributions from the $z$-dependent parts of the image self-interaction, $u_{\rm im}$. %The direct self-interaction, $u_0$, does not depend on coordinates, so it simply renormalizes the fugacity. 
Note that if there is no dielectric discontinuity, so that $u_\textrm{im}(\Av r,\Av r') = 0$, this term is identically equal to zero!  

%Taking all this into account, the strong-coupling interaction free energy for multipolar counterions is $\beta {\cal F}_\textrm{SC}=-\ln\,{\cal Z}_{\rm SC}$ and the part of the free energy that depends on the inter-surface separation 
%can be written in dimensionless form as
%\begin{equation}
%\frac{\beta {\cal F}_\textrm{SC}}{N}=\frac{D}{\mu}-2\,\ln\int_{- a}^{ a}\rmd z\int_{-1}^1 \rmd x\,\exp\bigl(-  w(z,x)\bigr),
%\label{FreeEn}
%\end{equation}
%where $w(z,x)$ is strictly due only to 
It then turns out that multipolar contributions (beyond monopoles) in the SC free energy, $\beta {\cal F}_\textrm{SC}=-\ln\,{\cal Z}_{\rm SC}$, enter only via self-image interactions  (see Ref. \cite{multipoles}). Therefore, {\em if} there is no dielectric discontinuity, the multipolar effects in this limit vanish!  Note that we have again assumed that the blip function $\Omega$ does not depend on the coordinate $z$, thus disregarding the possible entropic effects due to the finite size and anisotropy in the shape of counterions \cite{bohinc,woon}, which would be
relevant only for small inter-surface separation on the order of the counterion size, where other competing interactions, besides electrostatic, come into play as well \cite{Ben-Yaakov2011b,Conway1981}.
%. These entropic contributions to the partition function are relevant only for inter-surface separation on the order of the size of the counterion or smaller, where other types of interactions, besides electrostatic, come into play \cite{Ben-Yaakov2011b,Conway1981}.

\begin{figure*}[t]\begin{center}
\begin{minipage}[b]{0.33\textwidth}\begin{center}
\includegraphics[width=\textwidth]{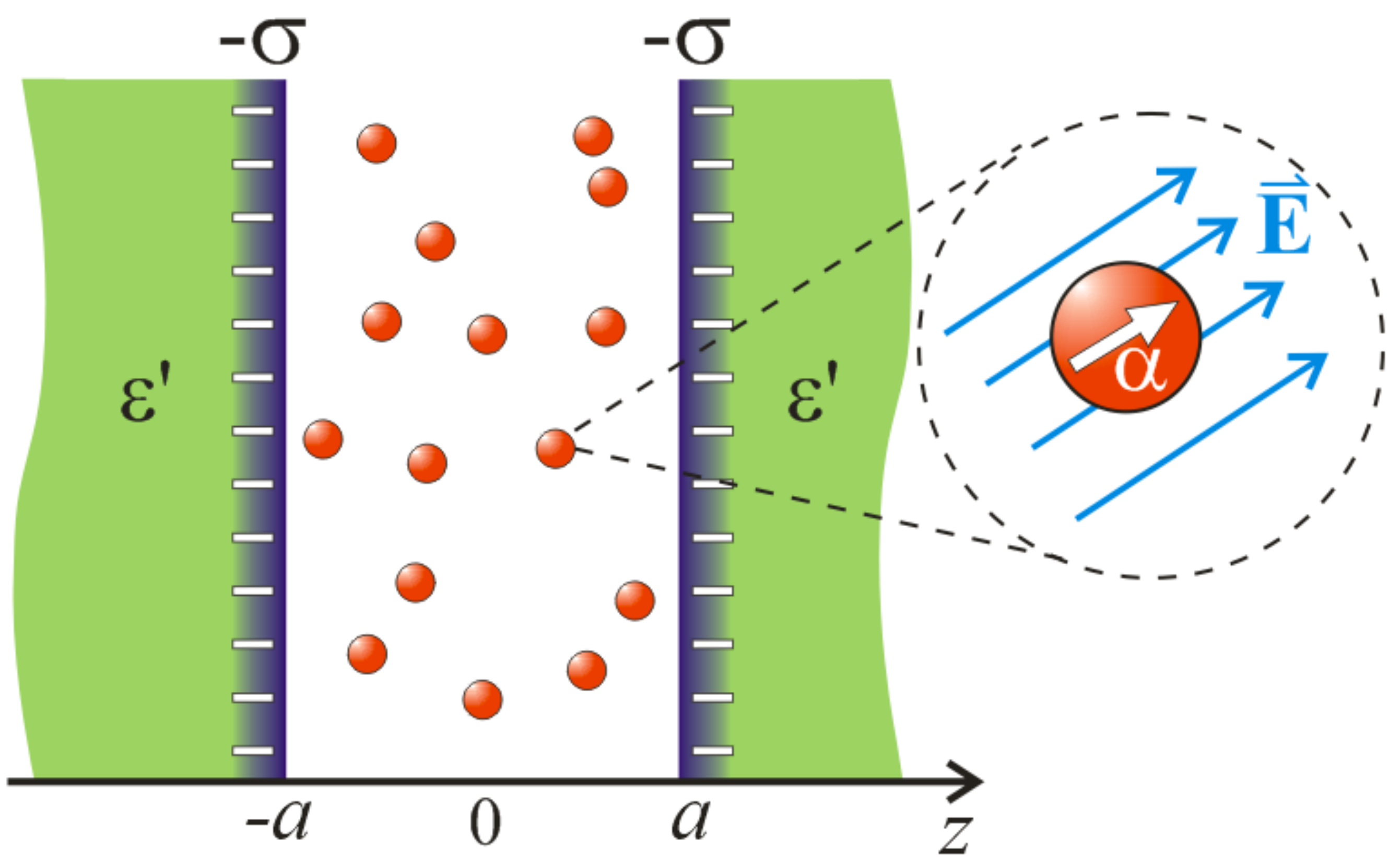} (a)
\end{center}\end{minipage}\hskip0.1cm
\begin{minipage}[b]{0.30\textwidth}\begin{center}
\includegraphics[width=\textwidth]{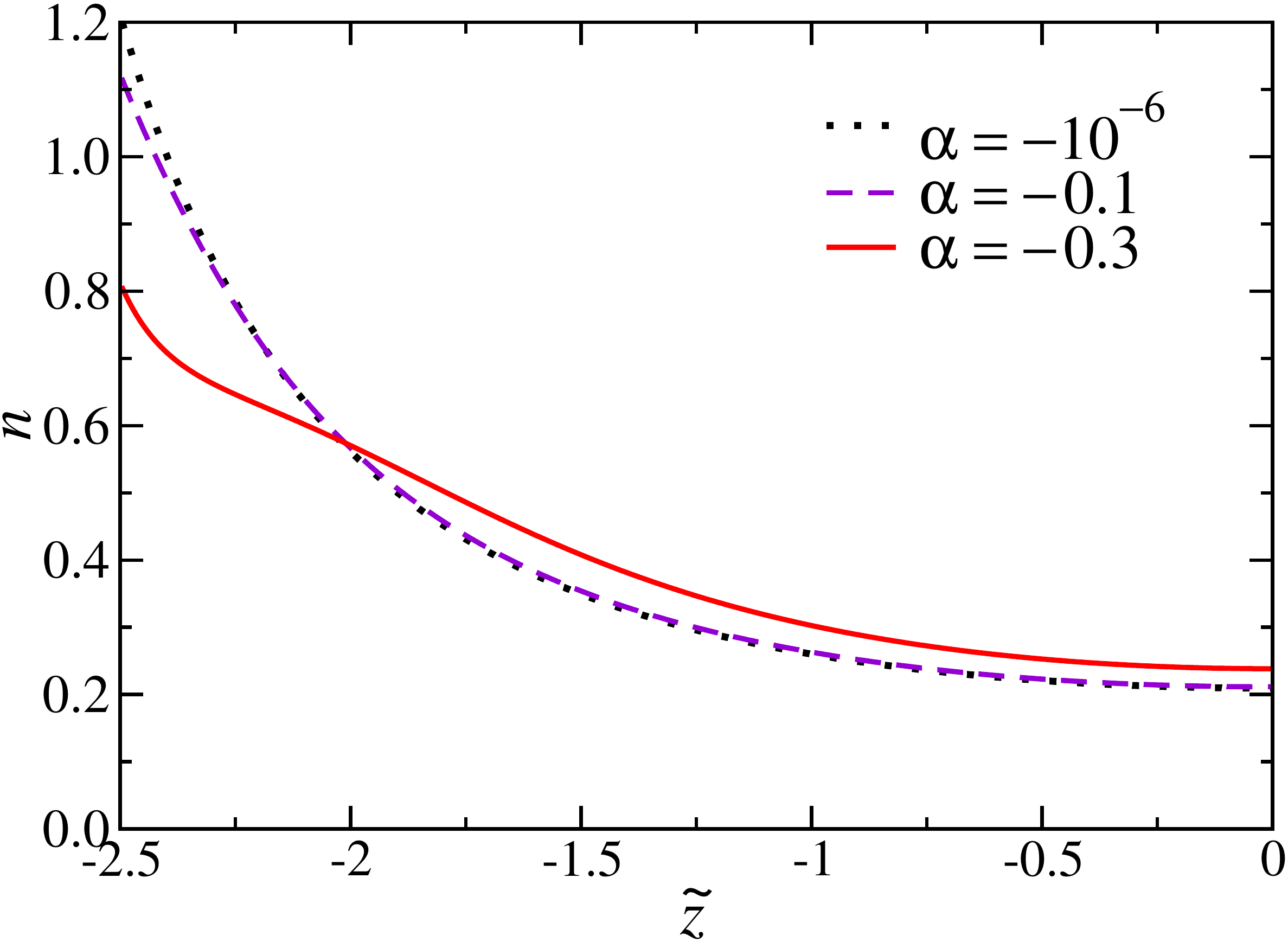} (b)
\end{center}\end{minipage}\hskip0.1cm
\begin{minipage}[b]{0.30\textwidth}\begin{center}
\includegraphics[width=\textwidth]{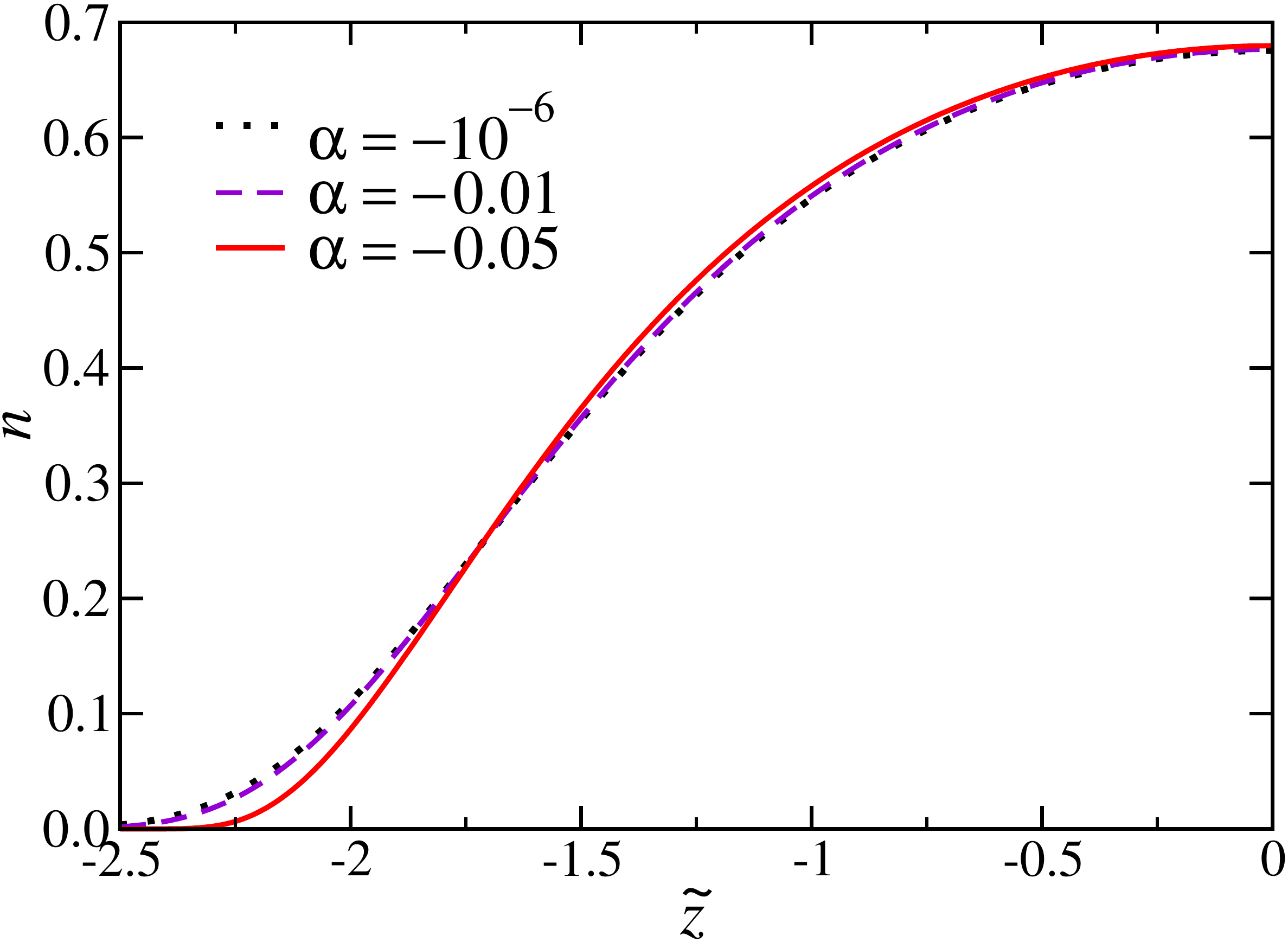} (c)
\end{center}\end{minipage}\hskip0.1cm
\caption{(a) Counterion polarizability, $\alpha$, introduces {\em effective interactions} between the counterions and the bounding surfaces. For negative excess polarizability, these additional interactions seem to be strongly repulsive and long ranged. 
(b) The WC counterions density profile (in units of $2\pi \ell_{\mathrm{B}} \sigma^2$) as a function of the rescaled position, $\tilde z=z/\mu$, 
in the slit is shown next to the left surface for rescaled inter-surface half-distance $\tilde a = a/\mu=2.5$ and different polarizability values $\alpha$ as indicated on the graph; here, 
the contribution from  fluctuations around the mean-field solution are taken into account as well. 
(c) Same as (b), but here we show the SC  density profile  
%for $R=1$, $\epsilon\ind{ext}=0.05$, $\Xi=10$  
for different $\alpha$. A clear effective repulsion is seen. Adapted from Ref. \cite{Demery}. }
\label{fig:polar}
\end{center}\end{figure*}

As for the inter-surface interaction pressure in the SC limit, the quadrupolar contribution can become negative (attractive) due to dielectric image contributions. On increase of the quadrupolar moment, the interaction becomes smaller and is eventually overwhelmed by repulsive contributions. The orientational order parameter in the SC limit indicates that the counterions are preferably aligned perpendicular to the $z$-axis, {\em parallel} with the surfaces (see Fig.~\ref{fig:rods}, bottom, left), which is contrary to the WC case. This is caused by strong quadrupole-quadrupole image repulsion when there is a  dielectric discontinuity at the surfaces; otherwise, quadrupolar effects are nil as discussed above. 
Therefore, in the plane-parallel geometry, higher-order multipoles in the internal structure of mobile ions can play an important role, but {\em only} for a dielectrically inhomogeneous system, i.e., when the bounding surfaces and the slit region possess different dielectric constants. Although, we should  also emphasize that multipolar effects can be present if the charged surfaces are curved (and thus generate spatially varying external potentials) {\em even} without dielectric discontinuities, but these effects remain largely unexplored. 

\subsection{Polarizable ions---a two-parameter model}

Apart from possible multipolar moments in addition to the monopolar charge, 
%that make a contribution to the free energy only in the case of dielectric discontinuities, 
we have been up to now dealing with a \emph{one-parameter model} of  mobile (counter-)ions, as by assumption they differ only in the amount of charge they bear.  Equally charged ions, such as $\rm Na^{+}$ and $\rm Li^{+}$, or  $\rm Ca^{2+}$ and $\rm Mn^{2+}$, are thus indistinguishable and have no chemical identity. One possible generalization of this model is  to additionally characterize the ion with its excess static ionic polarizability \cite{Ben-Yaakov2011,Ben-Yaakov2011b,Frydel2011,Sahin2,Sahin3} (see Fig.~\ref{fig:polar}a), proportional to the volume of the cavity created by the ion in the solvent. Static excess ionic polarizability is then a second parameter that differentiates between different, but equally charged ionic species, and thus obviously introduces ionic specificity into the theory.  Excess ionic polarizability studies go all the way back to the classical work by Debye on polar molecules \cite{Debye}.

For a system composed of polarizable monopolar counterions, the field self-interaction can be derived   \cite{Demery} in the form of Eq.~(\ref{fieldaction}), but with 
\begin{equation}
{\cal U}(\phi({\bf r}))=  \lambda \Omega({\mathbf r})\, \exp\left[{-\beta\frac{\alpha}{2}(\nabla\phi({\bf r}))^2-\rmi\beta qe_0\phi({\bf r})}\right],
\label{xi}
\end{equation}
where $\alpha$ is the excess polarizability of the counterions, while other details of the model are the same as for a standard counterion-only system. The excess polarizability is defined precisely as the difference between the aqueous solvent polarizability and the proper ionic polarizability, and may thus be negative \cite{Debye,Ben-Yaakov2011b}.

One should note here that the partition function depends on two parameters: the coupling constant $\Xi$ as well as the polarizability $\alpha$, i.e., it is a {\em two-parameter function}. These two parameters can be introduced in the following way:  we first rewrite the electrostatic coupling constant (Eq.~(\ref{aclcadfls})) in the form $\Xi = q^2 \ell_{\mathrm{B}}/\mu = 2\pi q^3 \ell_{\mathrm{B}}^2|\sigma| \equiv q^3 \Xi_0$, where we specifically decomposed the coupling parameter into its $q$ and $\sigma$ dependence. Then we introduce a rescaled polarizability that represents an additional independent parameter of the theory and is defined as $\tilde \alpha =  \alpha\left({\beta }/{(\beta q e_0 \mu)^2}\right)$. Also, instead of using the excess polarizability, we can use the {\em dielectric decrement} $\tilde\beta$ in units of inverse mole per liter (denoted by $\rm M^{-1}$) \cite{Ben-Yaakov2011}, i.e., $\alpha = \varepsilon_0 \tilde\beta$. The dielectric decrement for various salts is typically negative and on the order of $10~\rm M^{-1}$ in magnitude. 

Again, we introduce the corresponding WC-SC proxies as the saddle-point and the single-particle theory, based on the lowest-order virial expansion. 
Contrary to the cases described before, there exist at present no pertinent simulations that could help in assessing the accuracy of these approximations (however, see Ref.~\cite{Holmsimul}).

%The dimensionless polarizability represents an additional independent parameter of the theory. Finally we define the dimensionless fugacity as $ \tilde \lambda=2\pi\Xi \mu^3\lambda$. 

The saddle-point level, corresponding to the WC limit, can be derived in the form of a generalized PB equation governing the mean (real-valued) potential $\psi_\textrm{MF}=\rmi\phi$ as \cite{Ben-Yaakov2011b,Frydel2011}
\begin{equation}
\label{mf_bulk}
- \varepsilon_0\nabla\cdot\left[\left({\varepsilon} + \alpha n_\textrm{MF}({\mathbf{r}})\right)\nabla\psi_\textrm{MF}({\mathbf{r}})\right]= \rho_0({\mathbf{r}}) + n_\textrm{MF}({\mathbf{r}}),
\end{equation} 
where the mean-field counterion density is given by
\begin{equation}\label{def_n}
n_\textrm{MF}({\mathbf{r}})= \lambda \exp\left[\beta\frac{\alpha}{2}(\nabla\psi_\textrm{MF}({\mathbf{r}}))^2-\beta qe_0\psi_\textrm{MF}({\mathbf{r}})\right].
\end{equation}
Comparing this equation with Eq.~(\ref{eq:PB}), it is clear that the dielectric response function has a term proportional to the concentration of the ions, stemming from their dielectric decrement. This was in fact first noted by Bikerman in 1942 ~\cite{Bikerman,Hatlo2012}. The solutions of this equation have been investigated in great detail in Refs. \cite{Ben-Yaakov2011b,Frydel2011} and in general depend on the magnitude and sign of the excess polarizability. 

From numerical solutions of the above PB equation, it appears that the counterion polarizability introduces {\em effective interactions} between the ions and the bounding surfaces. For negative excess polarizability, these additional interactions seem to be strongly repulsive and long ranged. They again lead to a depletion of the ions in the vicinity of the surface \cite{Monica}, see Fig.~\ref{fig:polar}b. In the opposite case, the interactions are also repulsive, but weak, and the perturbations introduced by polarization are negligible. This yields a good measure for ion specificity, as the ions can be differentiated according to the sign of their polarizability.

%In the WC  limit, if ${\varepsilon} + \alpha n_\textrm{MF}$ becomes negative, a divergence appears for the fluctuations about the mean field and the system becomes unstable. This particular instability shows that for dense systems, the linear relation between concentration and dielectric decrement breaks down. Other non-linear solvation related effects \cite{Conway1981} that have not been taken into account would then take over and stabilize the system.

On the single-particle level, corresponding to the SC limit, ${\mathcal Z}_{\rm SC}^{(0)}$ is again given by the first term in Eq.~(\ref{partfun}), involving only electrostatic interactions between fixed  surface charges. The polarizability of the mobile ions appears in the first-virial-order partition function, which is again formally of a single-particle type and can be cast into the form \cite{Demery}
\begin{eqnarray}
{\cal Z}_{\rm SC}^{(1)} =  \int\, \rmd \vct r_0\,{z_{\rm SC}^{(1)}(\vct r_0)},
\end{eqnarray}
where we have introduced 
\begin{widetext}
\begin{equation}\label{z1}
{z_{\rm SC}^{(1)}(\vct r_0)}/{{\mathcal Z}_{\rm SC}^{(0)}} = \det\left(1+\alpha\nabla_i\nabla'_j u(\vct r_0,\vct r_0)\right)^{-1/2} \times \exp\left(\frac{1}{2} \nabla_i C'(\vct r_0)^T\left(\frac{1}{\alpha}+\nabla_i\nabla'_j u(\vct r_0,\vct r_0)\right)^{-1} \nabla_i C'(\vct r_0)-\frac{C'(\vct r_0)}{2}\right),
\end{equation}
\end{widetext}
with
$
%\kern-20pt {\cal A}_{ij}(\vct r_0) & = & \nabla_i\nabla'_j u(\vct r_0,\vct r_0),\label{def_A} \nonumber\\
C'(\vct r_0)  =  u(\vct r_0,\vct r_0)-\!\!\int \rho_0(\vct r)u(\vct r_0,\vct r)\,\rmd\vct r .
%C(\vct r_0) & = & C'(\vct r_0,\vct r_0)+\int \rho_0(\vct r)\rho_0(\vct r')u(\vct r,\vct r')d'\vct r d'\vct r' .\phantom{x}
$
Here, $u(\vct r,\vct r')$ is again just the Coulomb interaction potential and $\nabla$ as well as $\nabla'$ denote  the gradients with respect to the first and the second variable. 

Though the above form of the single-particle partition function appears to be quite complicated, it can be seen  straightforwardly that the first term in Eq.~(\ref{z1}) describes the thermal Casimir or zero-frequency vdW interaction between a single polarizable particle and the dielectric interfaces in the system. It can be written as
%\begin{widetext}
\begin{equation}
%{\textstyle\frac12} {\rm Tr }\log{(1+\alpha{\cal A}_{ij}(\vct r_0))} &=& 
{\textstyle\frac12} {\rm Tr}\,\ln{\left(1+\alpha \nabla_i\nabla'_j u(\vct r_0,\vct r_0)\right)} \simeq {\textstyle\frac12} \alpha~ {\rm Tr}\left[\nabla_i\nabla'_j u(\vct r_0,\vct r_0)\right].
\end{equation}
%\end{widetext}
For large $|\vct r_0|$, we obtain the scaling form $|\vct r_0|^{-3}$ corresponding to the zero-frequency vdW interaction between the polarizable particle and a single dielectric discontinuity \cite{Parsegian2005,Ninham1997}. 
%Due to the polarizability of the ions, it is also clear that the theory breaks down if the parameters are too extreme. 

The corresponding single-particle-level counterion density profile, Fig.~\ref{fig:polar}c,  exhibits strong image-repulsive interactions that deplete the vicinal space next to the bounding surfaces to an extent much larger than in the case of the saddle-point limit in Fig.~\ref{fig:polar}b \cite{Monica}.  Furthermore, on the single-particle level,  the effective permittivity around the ion may turn overall negative, leading to a field instability that shows up in the partition function. Other non-linear solvation-related effects~\cite{Conway1981} that have not been taken into account would then take over and stabilize the system.

%Though formally the field theory arising from our model is independent of the ionic size, it can be ascertained that on the single-particle level the partition function exhibits an ultra-violet or short-distance divergence~\cite{Demery}. 
%This divergence is associated directly with ionic polarizability, as it does not arise for non-polarizable ions. One can argue that the length scale used to cut-off ultra-violet divergences is thus the size of the polarizable ions, which is in line with Bikerman~\cite{Bikerman}, who long ago argued for the role of the ionic size.

In summary, for polarizable ions, the validity of the SC vs.\ WC description no longer depends on a single coupling parameter, but actually on two parameters. The parameter space is thus quite complicated and the validity of the WC-SC dichotomy is difficult to assess in general. The general conclusion would be that the contribution of polarizable counterions to the total partition function is highly non-additive at the weak coupling level, whereas it can sometimes be reduced to an additive contribution in the free energy at the strong coupling level, but only if the polarizability is large enough. Simply adding a vdW ion-polarizability-dependent contribution to the electrostatic potential of mean force is questionable \cite{Ninham1997,LoNostro}. 

\section{Conclusions}

We have provided a {\em guided personalized tour} of recent advances in Coulomb fluids based on the functional integral representation of the partition function via a field-action, as pioneered in the fundamental work of Edwards and Lenard \cite{Edwards}. This representation was taken later as a point of departure for the introduction of the WC-SC dichotomy in the description of a (primitive) counterion-only system bounded by two charged surfaces. The WC limit was shown to stem from the saddle-point description of the field-action, while the SC description was based on the lowest-order virial expansion. In the case of the primitive counterion-only model, the two limits can be derived directly and systematically. For more complicated, but also more realistic cases that imply a multiplicity of different length scales, one can identify the principal coupling parameter and introduce the pertaining saddle-point and the single-particle virial description as the proxies for the proper WC and SC limits.
The saddle-point and the single-particle virial form of the partition function can then be invoked for any system, irrespective of the number of parameters or coupling constants describing it, and thus present a convenient point of departure to derive powerful limiting laws even for much more complicated systems than envisioned in the original framework of the WC-SC dichotomy. 
We tried to make a clear case that this philosophy can provide a solid foundation and a fairly accurate guide for an approximate and sometimes even analytical treatment of Coulomb systems that formally do not easily yield themselves to a simple single-coupling-parameter description.

\section{Acknowledgment} 

R.P. acknowledges support by the U.S. Department of Energy, Office of Basic Energy Sciences, Division of Materials Sciences and Engineering under Award DE-SC0008176 in preparing this review.

\end{document}